\documentclass[12pt]{iopart}
\usepackage{tabulary,graphicx}
\usepackage{autonum}
\begin{document}
\title[]{Rare decays of $B_d$ mesons into four charged leptons in the Standard Model}

\author{A. V. Danilina$^{1,2}$, N. V. Nikitin$^{1, 2,3,4}$\footnote{Present address:
Federal State Budget Educational Institution of Higher Education M. V. Lomonosov Moscow State University, Skobeltsyn Institute of Nuclear Physics (SINP MSU), 1(2), Leninskie gory, GSP-1, Moscow 119991, Russian Federation}}

\address{$^1$Skobeltsyn Institute of Nuclear Physics, Moscow, Russia }
\address{$^2$NRC “Kurchatov Institute”  – ITEP, Moscow, Russia}
\address{$^3$Lomonosov Moscow State University, Physics Faculty, Moscow, Russia}
\address{$^4$The Moscow Institute of Physics and Technology, Dolgoprudny, Moscow Region, Russia}

\ead{anna.danilina@cern.ch, Nikolai.Nikitine@cern.ch}

\begin{abstract}
We present first theoretical predictions for various observables related to the $\bar B_d \to \mu^+ \mu^- e^+ e^-$ decay within the framework of the Standard Model. Our study encompasses the branching ratios, differential distributions, and forward -- backward leptonic asymmetries. To obtain these predictions, we account for several contributions: resonance contribution from $\rho(770)$ and significant contribution $\omega(782)$;  contributions from four charmonium resonances: $\psi(3770)$, $\psi(4040)$, $\psi(4160)$, and $\psi(4415)$; additional contributions from the “tails” of $J/\psi$ and $\psi(2S)$ resonances; non-resonant contributions arising from virtual photon emission  by a $b$ -- quark of $B_d$ meson, bremsstrahlung, and weak annihilation. In our calculations we employ the model of vector meson dominance (VMD) to assess the resonance contributions accurately.
We provide the predictions for the branching ratio of the $\bar B_d \to \mu^+ \mu^- e^+ e^-$ decay both with and without the resonant contribution from $\omega(782)$.
\end{abstract}

\section{Introduction}
In the realm of high-energy physics the quest to identify deviations from Standard Model (SM) predictions is a formidable challenge. While the SM has been proven remarkably successful, certain experimental results remain puzzling and unaccounted for within its framework. One prominent example is the indication of the existence of dark matter obtained from astrophysical experiments~\cite{DarkMatter}. The SM lacks fundamental particles with properties matching those of dark matter, making them potent argument in favor of new physics.

In addition to dark matter indications, various experimental hints have surfaced, suggesting deviations from SM predictions. For instance, the measurements of the muon's anomalous magnetic moment deviating from SM expectations~\cite{AnMu}. The intriguing discrepancies between experimental and theoretical values for the partial widths of $\bar B_{d, s} \to \mu^+\mu^-$ decays have been noted ~\cite{CMS:2014xfa,Aaboud:2016ire,Aaij:2017vad,Aaij:2021, Bmumucms, Aaij:2022}.

The modern high energy physics experiments such as BELLE~--~II~(KEK) and LHCb~(LHC, CERN) experiments show plan on collecting abundance of new data in heavy flavor physics field~\cite{BaBar:2012obs, LHCb:2015gmp, Belle:2015qfa, Belle}.
The expected data of high integrated luminosity would give an excellent opportunity to study rare processes.
It would give  access to extremely rare $B$~--~mesons decays suppressed in Standard Model, which make them powerful tools for scrutinizing
potential new physics effects. 

Hence, it becomes needful to obtain reliable theoretical predictions for such very rare $B$ -- meson decays and to identify the decay characteristics most suitable for detecting manifestations of beyond Standard Model (BSM) effects. Such attemptions hold the promise of shedding light on new physics phenomena that go beyond the established framework of the Standard Model.

In the Standard Model, rare four-leptonic decays of B -- mesons can be categorized into two types. There are decays of the first type, involving a significant number of electromagnetic and weak vertices at the tree level, leading to the final lepton state. Examples of such decays include $B^- \to \mu^- \mu^+ \mu^- \bar{\nu}_{\mu}$ and similar processes involving charged B and $B_c$ -- mesons. Decays of the second type are forbidden at the tree level and only occur at higher orders of perturbation theory through one-loop diagrams of the “penguin” and/or “box” type. These processes involve Flavor Changing Neutral Currents (FCNC) and are characterized by loop quantum diagrams, where the contribution of new virtual particles can be observable and measurable. An example of the second type is given by the processes $\bar{B}_{d,s} \to \mu^+ \mu^- e^+ e^-$ and $\bar{B}_{d,s} \to \mu^+ \mu^- \mu^+ \mu^-$.

At the LHC and the Belle II factory, both the first and second types of B - decays processes are subjects of investigation. Currently experimental studies have yielded upper limits for the partial widths of four-leptonic decays of $B_d$ and $B_s$ mesons~\cite{Aaij:2013lla},~\cite{Aaij:2016kfs},~\cite{4mu2021}. 
The theoretical investigation of rare leptonic decays of neutral $B^0_{d,s}$ -- mesons represents significant and compelling challenge for ongoing and future studies at the LHC and other experimental facilities. Predictions for these decays are limited to a  source~\cite{Dincer:2003zq}, which lacks consideration of resonant contributions. Additionally, rough estimation has been provided in~\cite{Danilina:2018uzr}. These existing predictions have their limitations and are insufficient for comprehensive understanding of these rare decays.

This paper focuses on investigating the characteristics of the $\bar B_d \to \mu^+ \mu^- e^+ e^-$ decay in the framework of the Standard Model. We conduct comprehensive calculation of the branching ratio for $\bar B_d \to \mu^+ \mu^- e^+ e^-$, as well as its differential characteristics. The theoretical analysis includes contributions from $d\,\bar{d}$ pairs, the contributions of charmonium resonances, non-resonance contributions from $b$ quark, leading contributions from weak annihilation, and bremsstrahlung effects.
By accounting for these various contributions, we aim to provide thorough and detailed understanding of the decay $\bar B_d \to \mu^+ \mu^- e^+ e^-$. 

\section{Effective Hamiltonian}
\label{sec:B2lllnuHeff}
The effective Hamiltonian describing Flavor Changing Neutral Currents ($b \to d$ ) is expressed in the framework of the Wilson expansion, as discussed in reference~\cite{BrMu}:
\begin{eqnarray}\label{eq:b2sll}
&&{ {\cal H}_{eff}^{b\to d\ell^{+}\ell^{-}}(x,\mu)\, =\, 
{\frac{G_{F}}{\sqrt2}}\, {\frac{\alpha_{em}}{2\pi}}\, V_{tb}V^*_{td}\, 
\left[
\,-2\, {\frac{C_{7\gamma}(\mu)}{q^2}}\, 
        \Big\{ m_b\,
 \left (\bar d\, i\sigma_{\mu\nu}\left (1+\gamma_5\right )q^{\nu}b\right )
        \nonumber \right .}\\
&&\qquad\qquad\quad {+\, m_d\, 
 \left (\bar d\, i\sigma_{\mu\nu}\left (1-\gamma_5\right )q^{\nu}b\right )
        \Big \}
\cdot\left ({\bar \ell}\gamma^{\mu}\ell \right )}\\
&&\qquad\qquad\quad{+\, 
{C_{9V}(\mu)}\left (\bar d  O_{\mu} b \right )
\cdot\left ({\bar \ell}\gamma^{\mu}\ell \right )\, +\, 
C_{10A}(\mu)\left (\bar d  O_{\mu} b \right )
\cdot\left ({\bar \ell}\gamma^{\mu}\gamma_{5}\ell \right ) \Big]},\nonumber 
\end{eqnarray}
 here, ${O_{\mu} = \gamma_\mu (I - \gamma^5)}$ and ${q^\nu}$ represents the four-momentum of the ${\ell^+\ell^-}$ pair;  $q^2 = q^{\nu}q_{\nu}$. The matrix $\gamma^5$ is defined as $\gamma^5 = i \gamma^0 \gamma^1 \gamma^2 \gamma^3$, with $\varepsilon^{0123} = -1$, and $\sigma_{\mu\,\nu} = \frac{i}{2}[\gamma_\mu,\gamma_\nu]$.

In this framework, the parameter $\mu = \textrm{5 GeV}$ functions as a scale demarcating the boundary between short and long-distance contributions from strong interactions. The Fermi constant is represented as $G_F$, while the Cabibbo-Kobayashi-Maskawa (CKM) matrix is characterized by the matrix elements $V_{td}$ and $V_{tb}$. Within this context the foundational constituents of the Standard Model comprise light degrees of freedom, namely $u, d, s, c,$ and $b$ quarks, along with leptons, photons, and gluons. 

Taking into account the heavy degree of freedom, namely the $W, Z,$ and $t$ quark, these aspects are embedded within the Wilson coefficients $C_i(\mu)$. Key criterion for these coefficients is set by the condition that $C_2(M_W) = -1$. The numerical values of these coefficients, specifically $C_i(\mu)$, are sourced from references \cite{MNK, MN2004}, wherein $\mu = \textrm{5 GeV}$.

Weak annihilation is particularly relevant when studying decays involving $B$-mesons, as it can significantly contribute to the angular distributions. The structure of the Hamiltonian describing weak annihilation takes the shape of:
\begin{eqnarray}
\label{WEAK}
{\cal H}_{eff}^{\bar{B_d}-Q\bar{Q}}(x)\, =\,- 
{\frac{G_{F}}{\sqrt2}}\,  V_{Qb}V^*_{Qd}\,a_1(\mu)(\bar{d}\,O^\mu\,b)(\bar{Q}\,O_\mu\,Q),
\end{eqnarray}
where $Q =\{u,c\}$ and $a_1(\mu = \textrm{5 GeV}) \approx - 0.13$ \cite{NeuSt}.

The Hamiltonian associated with electromagnetic interaction is expressed as:

\begin{eqnarray}\label{eq:belm}
{\cal H}_{em}(x)\, =\, -\, e\,\sum\limits_f\, Q_f
\left (
\bar f(x)\,\gamma^{\mu} f(x)
\right )\,
A_{\mu}(x)\, =\, -\, j_{em}^{\mu}\,A_{\mu}(x),
\end{eqnarray}
here, the electric charge $e = |e| > 0$ is scaled by $e^2 = 4 \pi \alpha_{em}$, with $\alpha_{em} \approx 1/137$ representing the fine structure constant, $Q_f$ signifies the charge of the fermion with flavor $f$ measured in units of the elementary charge $e$. The symbol $f(x)$ denotes the fermionic field associated with flavor $f$, while $A_{\mu}(x)$ corresponds to the electromagnetic four-potential.
\\

\section{Basic contributions to the $\bar B_d \to \mu^+ \mu^- e^+ e^- $ decay}
\label{s3}
\begin{figure}[ht!]
\begin{tabular}{c}
\includegraphics[width=0.55\linewidth]{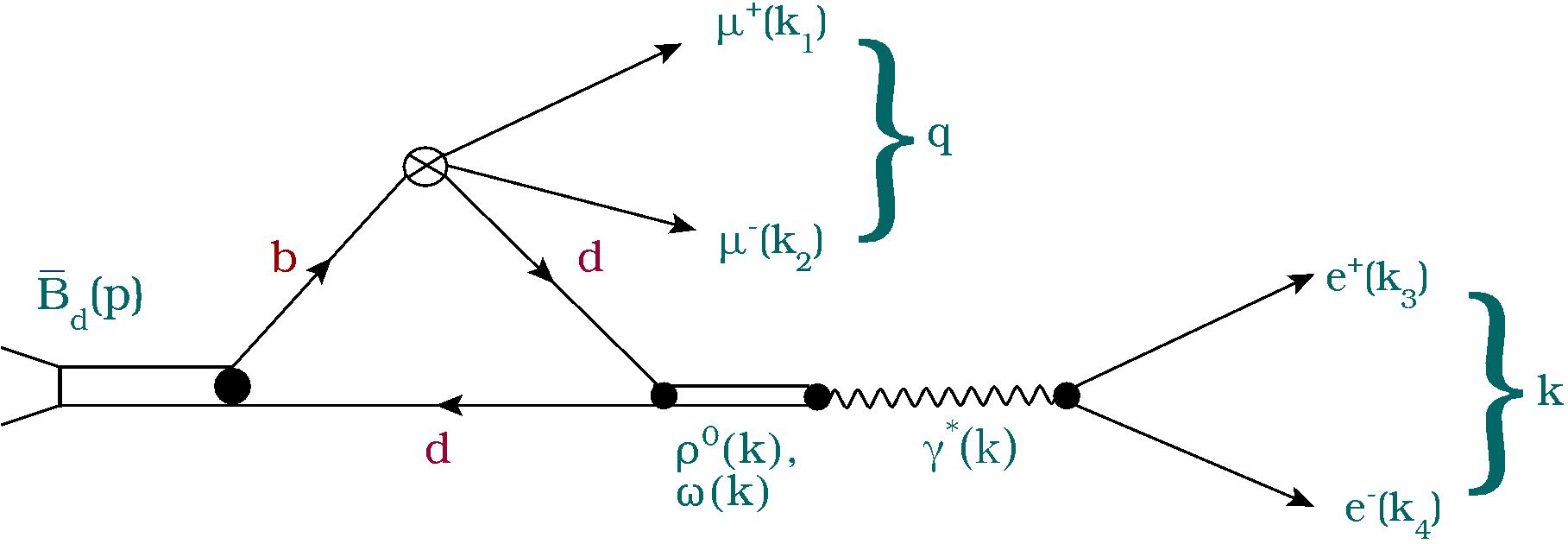}
\includegraphics[width=0.55\linewidth]{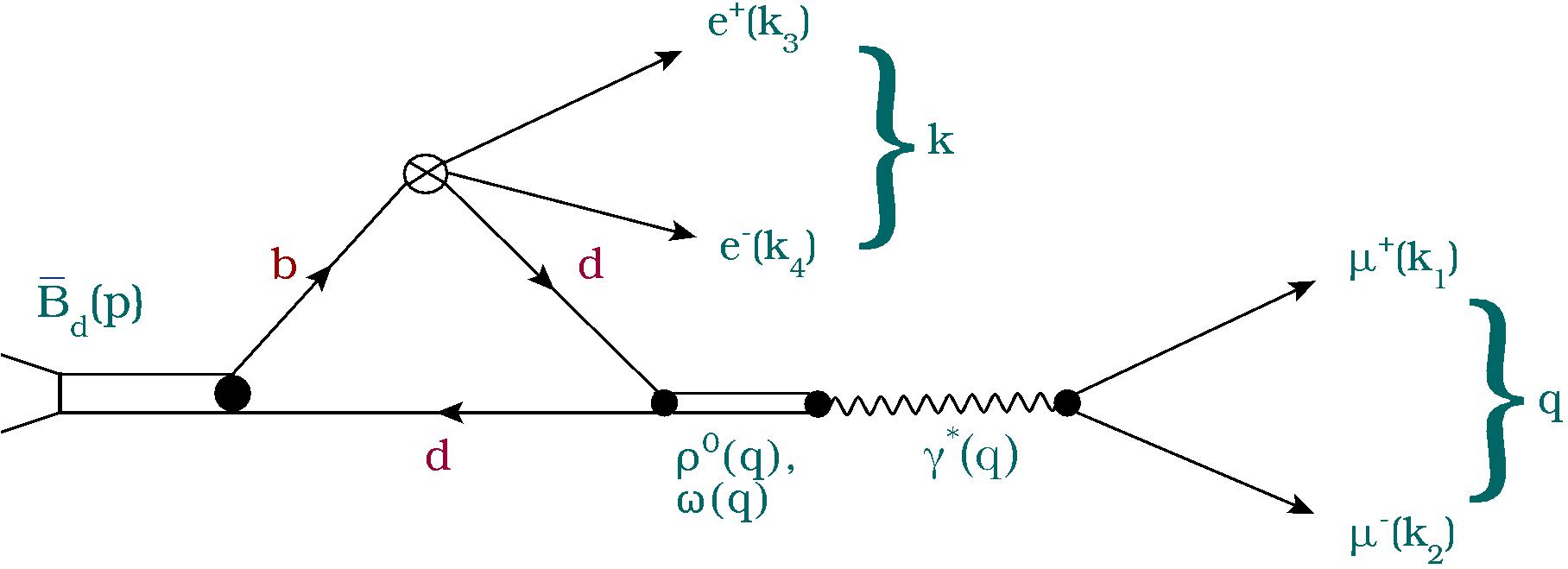} 
\end{tabular}
\caption{\label{fig:FRhoOmega} 
Emission diagram of a virtual photon by a $d$ -- quark of $B_d$ meson.}
\end{figure}

\begin{figure}[h!]
\begin{tabular}{c}
\includegraphics[width=0.55\linewidth]{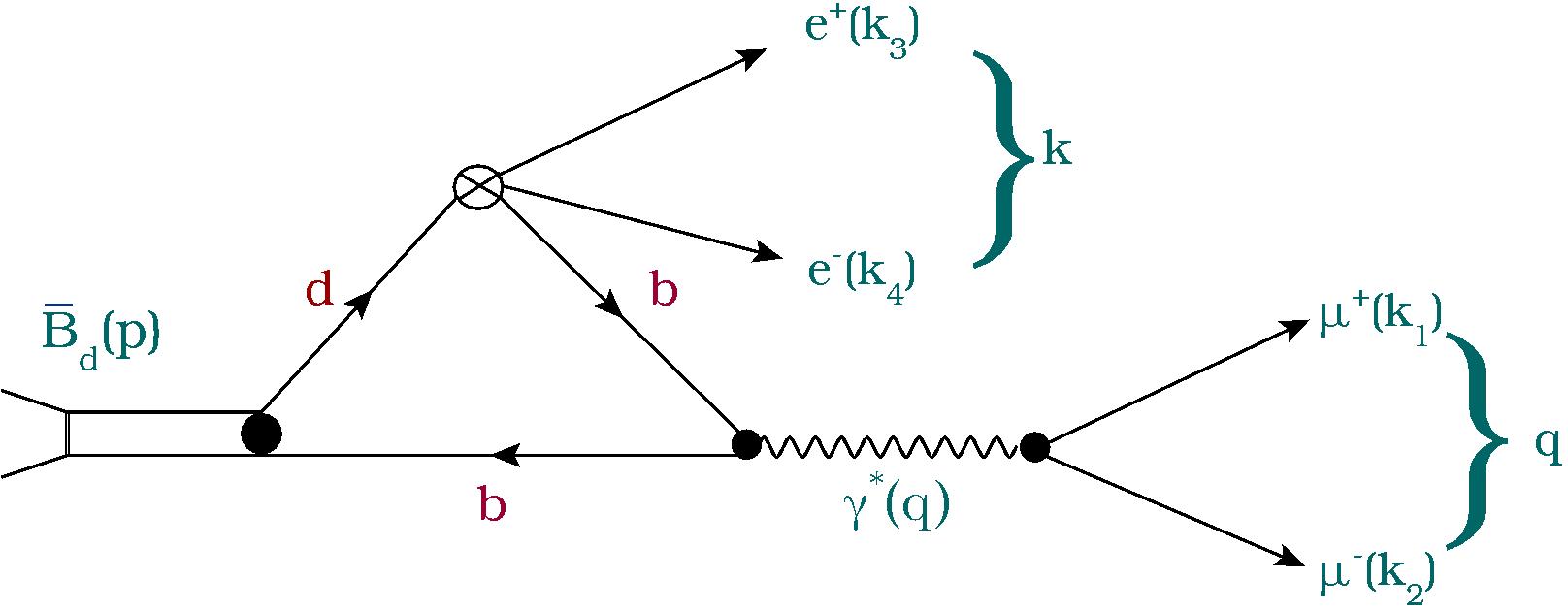}\,
\includegraphics[width=0.55\linewidth]{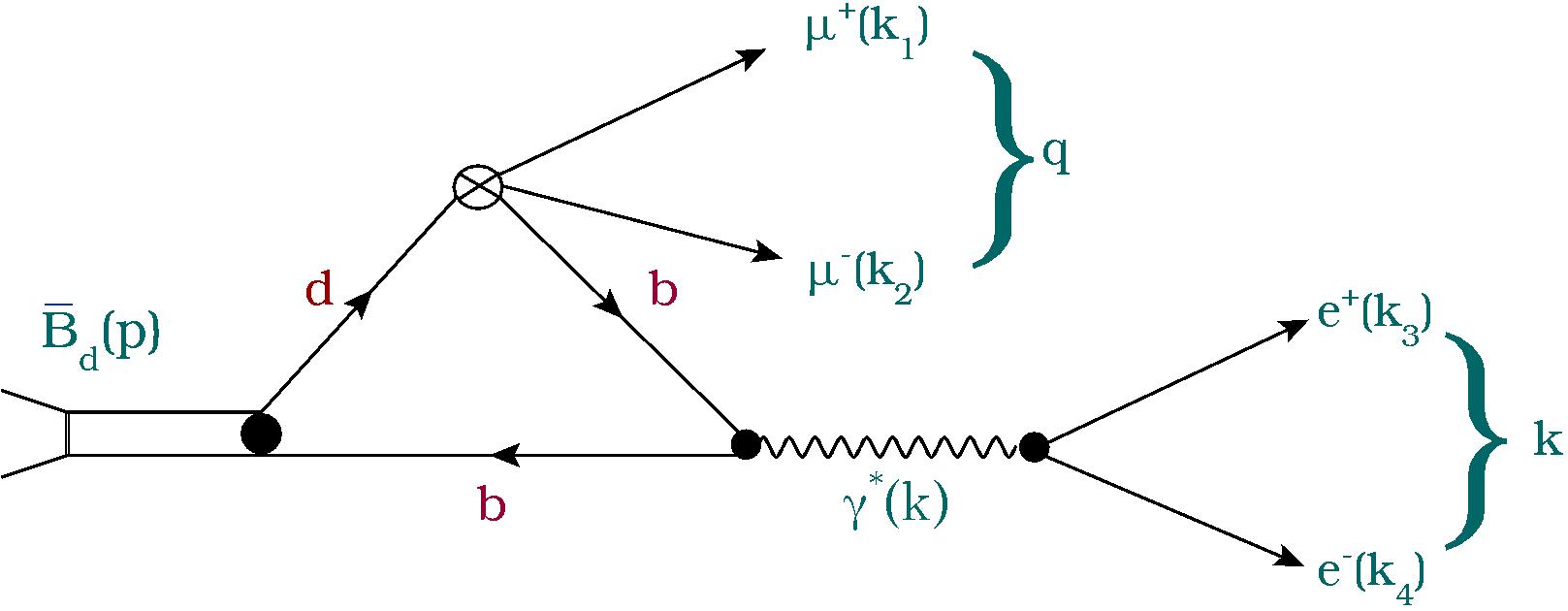} 
\end{tabular}
\caption{\label{fig:Fbbarb} 
Emission diagrams of a virtual photon by a $b$ -- quark of $B_d$ meson.}
\end{figure}

\begin{figure}[h!]
\begin{tabular}{cc}
\includegraphics[width=0.35\linewidth]{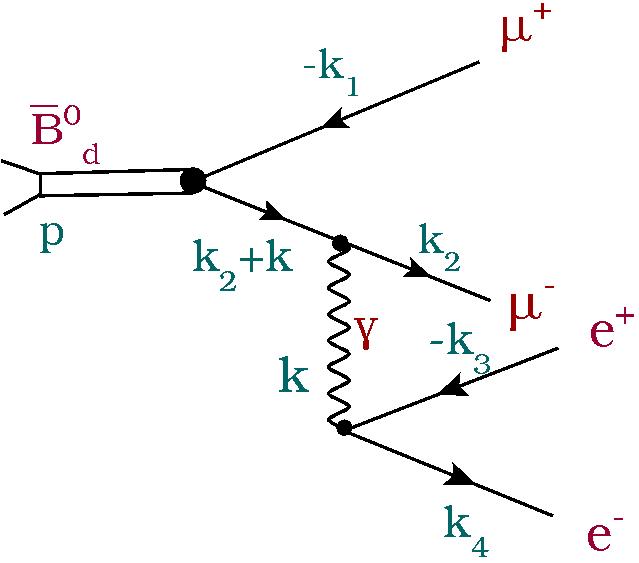} &\,\,\,\,\,\,\,\,\,\,\,\,\,\,\,\,\,\,\,\,\,\,\,\,\,\,\,\,\,\,\,\,\,\,\,\,\,\,\,\,\,\,\,\,\,
\includegraphics[width=0.35\linewidth]{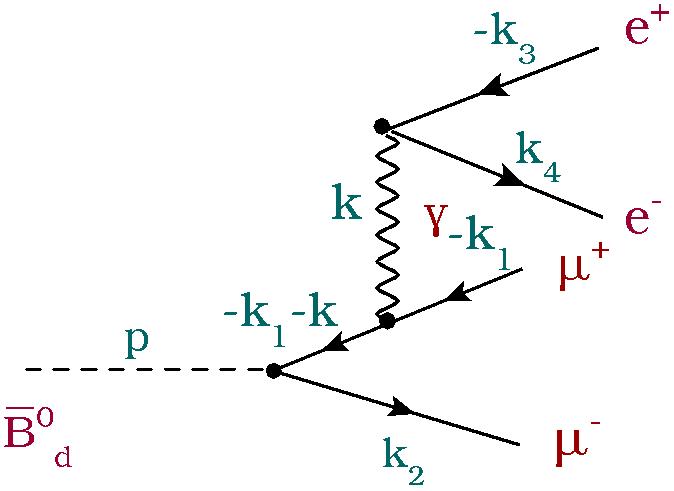} \\
\includegraphics[width=0.35\linewidth]{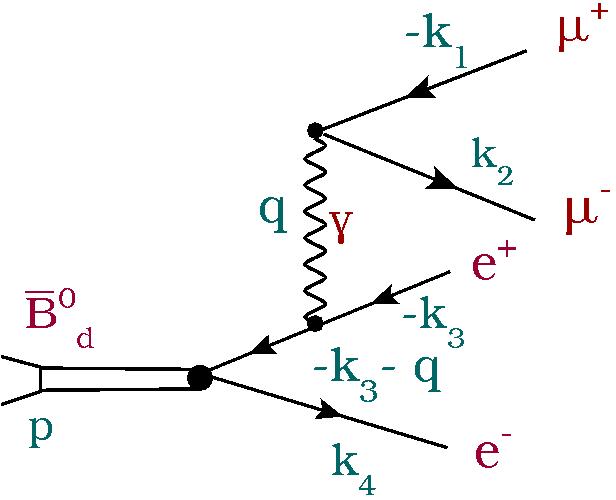} &\,\,\,\,\,\,\,\,\,\,\,\,\,\,\,\,\,\,\,\,\,\,\,\,\,\,\,\,\,\,\,\,\,\,\,\,\,\,\,\,\,\,\,\,\,
\includegraphics[width=0.35\linewidth]{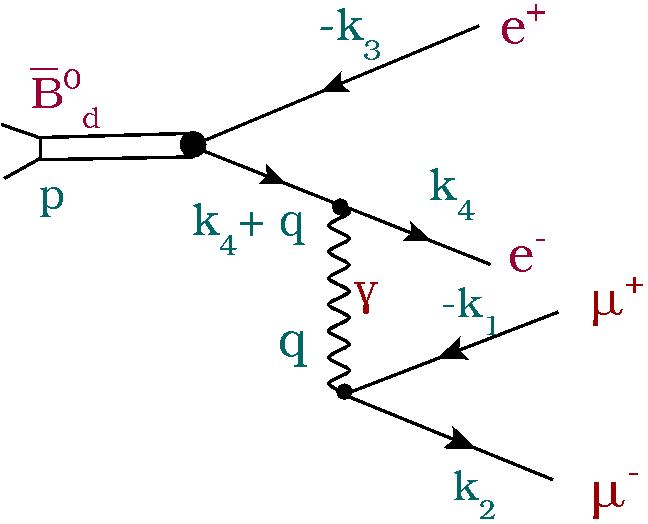} \\
\end{tabular}
\caption{\label{fig:Fbrem} 
Emission diagram of a virtual photon by leptons in the final state.}
\end{figure}

The comprehensive description of the $\bar B_d\to\mu^+(k_1)\mu^-(k_2)e^+(k_3)e^-(k_4)$ decays entails six primary types of contributions. The first type originates when a virtual photon is emitted from the $d$ -- quark, as illustrated in Fig.~\ref{fig:FRhoOmega}. The second type is associated with contribution stemming from photon, emitted from $b$ -- quark of $B_d$ -- meson. This contribution  depicted in Fig.~\ref{fig:Fbbarb}. The resonant contributions from $u\bar{u}$ and $c\bar{c}$ account for the third and fourth types, respectively. The fifth type pertains to the bremsstrahlung, where a virtual photon is emitted by one of the leptons in the final state, depicted in Fig.~\ref{fig:Fbrem}.  Lastly, the sixth type is linked to weak annihilation processes, as shown in Fig.~\ref{fig:WA}. The momenta $q = k_1 + k_2$ and $k = k_3 + k_4$ (refer to Appendix~\ref{fig:cinematic}).

\begin{figure}[t]
\begin{minipage}[h]{1.0\linewidth}
\center{\includegraphics[width=8cm]{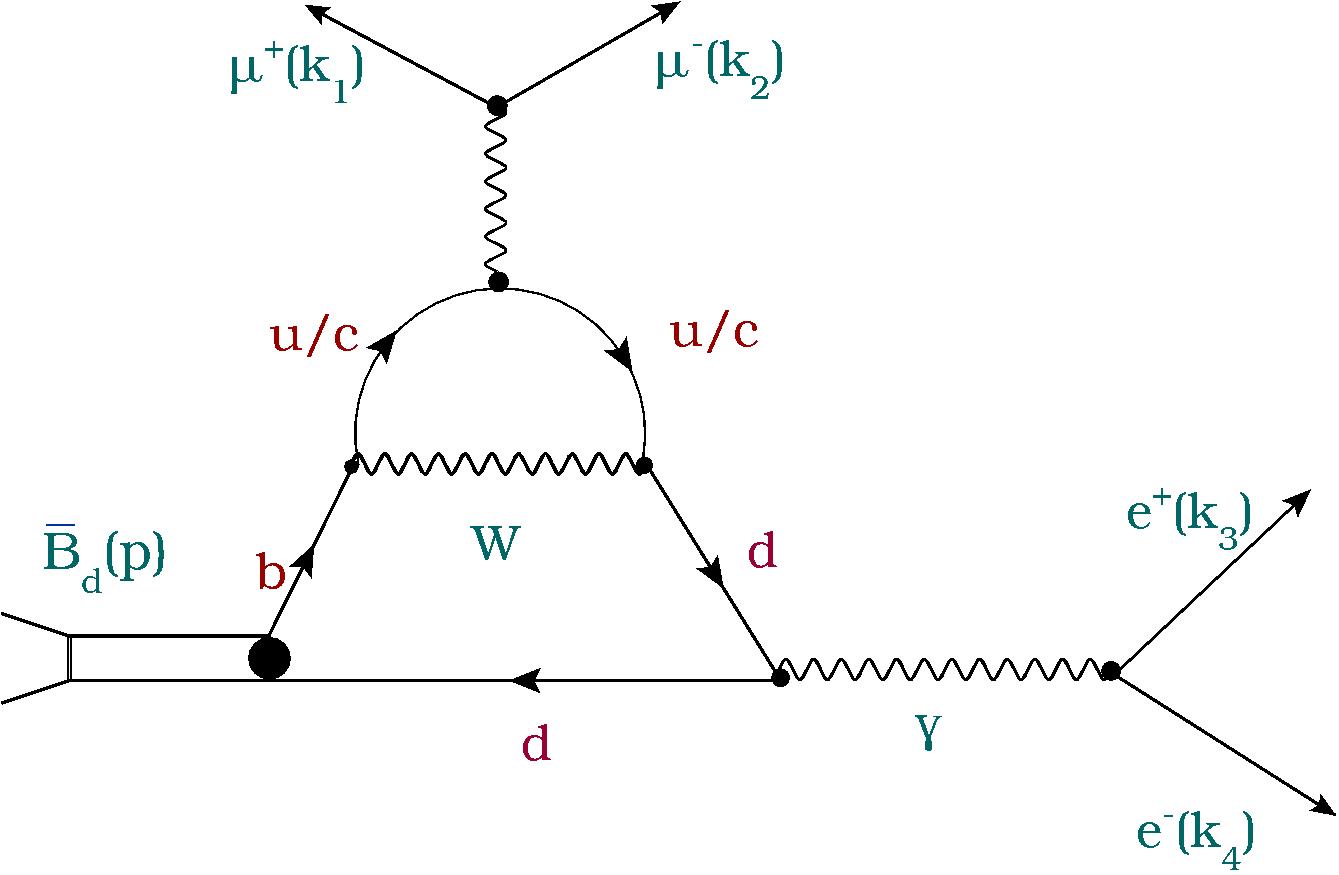}} \\
\end{minipage}
\caption{Ex. of diagram of the “penguin” type corresponding to loop contributions from $u/c$ quarks.}
\label{Bdpingiun}
\end{figure}

Utilizing the Hamiltonians (\ref{eq:b2sll}), (\ref{eq:belm}) and (\ref{WEAK}), we can deduce that the amplitude using the diagramms from Fig.~\ref{fig:FRhoOmega}, Fig.~\ref{fig:Fbbarb} and Fig.\ref{Bdpingiun} can be expressed as follows:

\begin{eqnarray}
\label{Amp}
\fl
{\cal M}_{fi}^{(1234)} = \sqrt2\,G_{F}\,\alpha_{em}^2\, V_{tb}V^*_{td}M_1\Bigg[\nonumber\\
 \hphantom{+}\fl\Bigg[-\frac{a^{(VV)}}{M^2_1}\,\varepsilon_{\mu\,\alpha\, k\, q}\, -\, i b^{(VV)}\, g_{\mu \alpha}\, +\, 2i\,\frac{c^{(VV)}}{M^2_1}\, q_{\alpha} k_{\mu}\,\Bigg]j^{\mu} (k_2,\, k_1)\, J^{\alpha} (k_4,\, k_3)\, +\nonumber\\
 \fl+\Bigg[-\frac{a^{(VA)}}{M^2_1}\,\varepsilon_{\mu\,\alpha\, k\, q}\, -\, i  b^{(VA)}\, g_{\mu \alpha}\, +\, 2i\,\frac{c^{(VA)}}{M^2_1}\, q_{\alpha} k_{\mu}\, +\,i\,\frac{g^{(VA)}}{M^2_1}\, k_{\mu} k_{\alpha}\,\Bigg]j^{\mu} (k_2,\, k_1)\, J^{\alpha\,5} (k_4,\, k_3)\,+\nonumber\\
 \fl+\Bigg[-\frac{a^{(AV)}}{M^2_1}\,\varepsilon_{\mu\,\alpha\, k\, q}\, -\, i  b^{(AV)}\, g_{\mu \alpha}\, +\, 2i\,\frac{c^{(AV)}}{M^2_1}\, q_{\alpha} k_{\mu}\,+\,i\,\frac{d^{(AV)}}{M^2_1}\, q_{\mu} k_{\alpha}\,\Bigg]j^{\mu\,5} (k_2,\, k_1)\, J^{\alpha} (k_4,\, k_3)\,+\nonumber\\
 \fl+\Bigg[-\frac{a^{(AA)}}{M^2_1}\,\varepsilon_{\mu\,\alpha\, k\, q}\, -\, i  b^{(AA)}\, g_{\mu \alpha}\, +\, 2i\,\frac{c^{(AA)}}{M^2_1}\, q_{\alpha} k_{\mu}\,+ ...\,\,\,\,\,\,\,\,\\ \,\,\,\,\,\,\,\,\,\,\,\,\,\,\,\,\,\,\,\,\,\,\,\,\,\,\,\,\,\,\,\,\,\,\,\,\,\,\,\,\,\,...+ i\,\frac{d^{(AA)}}{M^2_1}\, q_{\mu} k_{\alpha}\,+\,i\,\frac{g^{(AA)}}{M^2_1}\, k_{\mu} k_{\alpha}\,\Bigg]j^{\mu\,5} (k_2,\, k_1)\, J^{\alpha\,5} (k_4,\, k_3)\,\Bigg],\nonumber
\end{eqnarray}

In this context, $M_1$ corresponds to the mass of the $\bar B_d$ meson, and the currents are outlined as follows:
\begin{eqnarray}
j^{\mu} (k_2,\, k_1) = \bar\mu(k_2)\,\gamma^\mu\mu(-k_1),\,\, J^{\alpha} (k_4,\, k_3) = \bar e(k_4)\,\gamma^\alpha\,e(-k_3);\nonumber\\
j^{\mu\,5} (k_2,\, k_1) = \bar\mu(k_2)\,\gamma^\mu\,\gamma^5\,\mu(-k_1),\,\, J^{\alpha\,5} (k_4,\, k_3) = \bar e(k_4)\,\gamma^\alpha\,\gamma^5\,e(-k_3).\nonumber 
\end{eqnarray}

The functions are denoted by $a^{(IJ)} \equiv a^{(IJ)} (x_{12},\, x_{34})$, $b^{(IJ)} \equiv b^{(IJ)} (x_{12},\, x_{34})$, $c^{(IJ)} \equiv c^{(IJ)} (x_{12},\, x_{34})$, $d^{(IJ)} \equiv d^{(IJ)} (x_{12},\, x_{34})$, and $g^{(IJ)} \equiv g^{(IJ)} (x_{12},\, x_{34})$, where $x_{12} = \frac{q^2}{M^2_1}$  and  $x_{34} = \frac{k^2}{M^2_1}$. Here, the indices ${IJ\, =\,\{VV,\, VA,\, AV,\, AA\}}$ are precisely defined in section~\ref{sec;ABC}.

The contribution to the decay amplitude (\ref{Amp}) stemming from Fig.\ref{fig:FRhoOmega} can be computed employing the Vector Meson Dominance model (VMD). In this context, two scenarios are considered: virtual photon emission from intermediate vector meson giving rise to a $\mu^+\mu^-$ pair (depicted in Fig.\ref{fig:FRhoOmega}, left diagram), and virtual photon emission leading from intermediate vector meson  to an $e^+ e^-$ pair (depicted in the Fig.\ref{fig:FRhoOmega}, right diagram). In both cases the intermediate vector mesons $\rho(770)$ and $\omega(782)$ come into play, and their effects are accounted in the VMD framework. This is reflected in the coefficients $a^{(IJ)}$, $b^{(IJ)}$, $c^{(IJ)}$, $d^{(IJ)}$, and $g^{(IJ)}$ as discussed in ~\ref{sec;ABC}. Subsequent calculations reveal that the $\omega(782)$ resonance plays a main role in contributing to the amplitude of the $\bar B_d \to \mu^+ \mu^- e^+ e^-$ decay.

It is worth noting that when calculating contributions from $\rho^0$ and $\omega$ mesons, only the components of $\rho^0$ and $\omega$ mesons containing a $d\bar d$ pair are considered. Since the quark structure of $\rho^0$ and $\omega$ mesons is composed of linear combinations of $u\bar u$ and $d\bar d$ pairs, isotopic coefficients $I_i$ are introduced to isolate contributions solely from the $d\bar d$ pair. 
By definition,
\begin{equation}
    I_{\rho^0} = \langle \rho^0 | \bar d\, d \rangle = -\frac{1}{\sqrt{2}}, \quad I_{\omega} = \langle \omega | \bar d\, d \rangle = \frac{1}{\sqrt{2}}.
\end{equation}

Contributions stemming from charmonium vector resonances emerge from the effective Hamiltonian governing the transition ${b \to d\ell^{+}\ell^{-}}$ (\ref{eq:b2sll}) and are encompassed within coefficients $a^{(IJ)}$, $b^{(IJ)}$, and $c^{(IJ)}$.

The factorizable contributions originating from $c\bar{c}$ and $u\bar{u}$ components (see Fig. \ref{Bdpingiun}) of the amplitude can be described as an addition, dependent on the momentum transfer $q^2$ to the coefficient $C_{9V}$:
$$C_{9V}(\mu)\,\to\,
{C^{eff}_{9V}(\mu, q^2) \, =\,
C_{9V}(\mu)\, +}\,{\Delta C_{9V}^{c\bar c\, +\, u\bar u}(\mu, q^2)}.
\nonumber
$$

In this context, $C_{9V}(\mu)$ represents the Wilson coefficient ($C_{9V}(\mu = \textrm{5 GeV})=-4.21 $), while $\Delta C_{9V}^{c\bar c\, +\, u\bar u}(\mu, q^2)$ stands for the non-perturbative correction that encompasses both loop and resonant effects from $J/\psi$, $\psi(2S)$, $\rho^0(770)$, and $\omega(782)$~\cite{BrMu, KrSe}. This coefficient comprises fixed components that hinge on the $\mu$-scale, as well as contributions arising from loops involving $c\bar c$ and $u\bar u$ quarks, along with contributions from vector resonances.

The specific configuration of $C^{eff}_{9V}(\mu, q^2)$ can be delineated following the guidance provided in the reference \cite{MNS}.

In line with the experimental procedure established for excluding the contributions from $J/\psi$ and $\psi(2S)$ \cite{Aaij:2016kfs,Aaij:2013lla,4mu2021}, we consider only the residual effects from the $J/\psi$ and $\psi(2S)$ resonances during the theoretical computation of the differential characteristics and branching ratio for the $\bar B_d \to \mu^+ \mu^- e^+ e^-$ decay. 

The subsequent component contributing to the amplitude (\ref{Amp}) arises from the non-resonant Fig.~\ref{fig:Fbbarb}. This non-resonant contribution is reliant on variables $q^2$ and $k^2$, and its expressed through form factors (as defined in~\ref{sec;ABC}).

We consider the group of diagrams associated with bremsstrahlung processes, wherein a virtual photon is emitted by one of the leptons in the final state as depicted in Fig. \ref{fig:Fbrem}. This set comprises four distinct diagrams, each corresponding to photon emission by a lepton in the final state.

The amplitude for the ${\mu^+\mu^-}$-- pair emitted by electron and positron in the final state is:
\begin{eqnarray}
\label{MuBrem}
{{\cal M}_{fi}^{(\mu)}} &{=}&
{\sqrt{2}\, G_{\textrm{F}}\,\alpha^2_{em}\, V_{tb}\, V_{td}^*\, \Big (\overline{\mu}(k_2){\gamma^\mu}\mu(-k_1) \Big )\,
\Big [}
\nonumber\\
&&{ i\,{ d^{(VP)}} (x_{12},\, x_{123},\, x_{124})\, k_{\mu}\,
\Big (\overline{e}(k_4){\gamma^5} e(-k_3) \Big )\, +}
\\
&{+}&
{ f^{(VT)}} {(x_{12},\, x_{123},\, x_{124})\,\varepsilon_{\mu \nu \alpha \beta}\, p^{\nu}\,
\Big (\overline{e}(k_4){\gamma^\alpha \gamma^\beta} e(-k_3) \Big )
\Big]}.
\nonumber
\end{eqnarray}

Analogously, a closely resembling expression can be derived for the emission of the $e^+e^-$ pair by the muons present in the final state:
\begin{eqnarray}
\label{EBrem}
{{\cal M}_{fi}^{(e)}} &{=}&
{\sqrt{2}\, G_{\textrm{F}}\,\alpha^2_{em}\, V_{tb}\, V_{td}^*\, \Big (\overline{e}(k_2){\gamma^\mu}e(-k_1) \Big )\,
\Big [}
\nonumber\\
&&{ i\,{ d^{(VP)}} (x_{12},\, x_{123},\, x_{124})\, k_{\mu}\,
\Big (\overline{\mu}(k_4){\gamma^5} \mu(-k_3) \Big )\, +}
\\
&{+}&
{ f^{(VT)}} {(x_{12},\, x_{123},\, x_{124})\,\varepsilon_{\mu \nu \alpha \beta}\, p^{\nu}\,
\Big (\overline{\mu}(k_4){\gamma^\alpha \gamma^\beta} \mu(-k_3) \Big )
\Big]}.
\nonumber
\end{eqnarray}

To accurately compute the pole structure in bremsstrahlung processes, the presence of non-zero lepton masses needs to be accounted for. This adjustment is carried out in (\ref{MuBrem}).

The final contribution under our consideration pertains to the weak annihilation processes. These processes emerge from lowest-order diagrams that describe the contribution of $c\bar{c}$ and $u\bar{u}$ pairs after integrating out the degrees of freedom associated with the $W$ -- bosons. Thus, we incorporate the contribution from the axial anomaly, resulting in the subsequent structure for the amplitude:

\begin{figure}[h!]
\begin{center}
\begin{tabular}{c}
\includegraphics[width=0.5\linewidth]{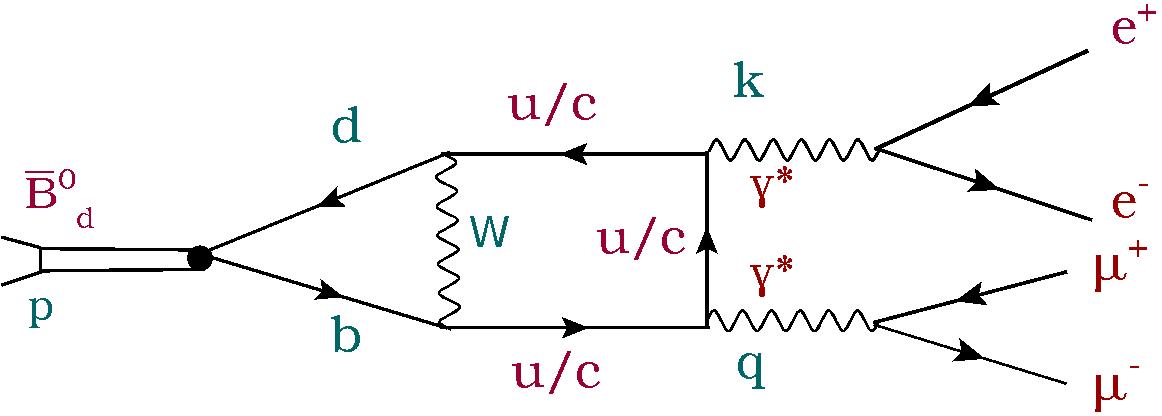}
\includegraphics[width=0.5\linewidth]{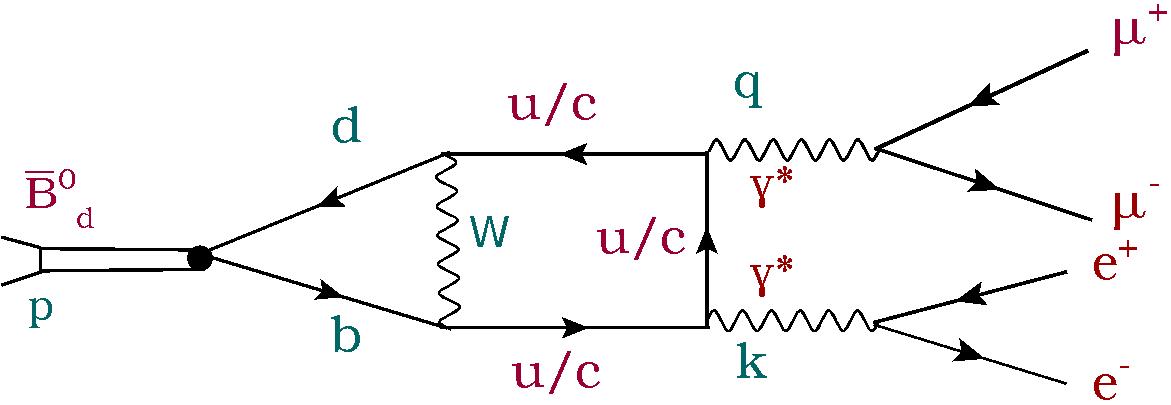} 
\end{tabular}
\caption{
The diagrams for the weak annihilation processes.}
\label{fig:WA}
\end{center}
\end{figure}

\begin{eqnarray}
\
&&{{\cal M}_{fi}^{(WA)}}\,{= \,\frac{32 \sqrt{2}}{3 \pi}\,\frac{G_{\textrm{F}}}{M^3_1}\,\alpha^2_{em}\, \left ( V_{ub}\, V_{ud}^*+V_{cb}\, V_{cd}^* \right )\, a_1 (\mu)\, {\hat f}_{B_d}} 
\nonumber \\
&&
{\frac{1}{x_{12}\, x_{34}}\,\varepsilon_{\mu\alpha k q}
\Big (\overline{\mu}(k_2)\,{\gamma^\mu}\mu(-k_1) \Big )\,
\Big (\overline{e}(k_4)\,{\gamma^\alpha} e(-k_3) \Big )}.
\end{eqnarray}
In the given equation, ${\hat f}_{B_d} = \frac{f_{B_d}}{M_1}$. For simplicity, we neglect the ${\left (m_c/M_1\right )^2}$ and ${\left (m_u/M_1\right )^2}$ corrections.

The weak annihilation contribution is magnified at lower four-momenta. This influence decreases by a factor related to the mass of the heavy quark, as compared to the contributions examined in the preceding sections.

\section{The calculation of the branching ratio for the $\bar B_d \to \mu^+ \mu^- e^+ e^- $ decay}

 To obtain characteristics of $B_d$ meson decays, it is essential to have a theoretically grounded modeling approach. To achieve this, we have developed a Monte Carlo model for the EvtGen package~\cite{Evt}. This model ensures the accurate simulation of the $\bar B_d \to \mu^+ \mu^- e^+ e^- $ decay, encompassing all the theoretical contributions discussed in Section~\ref{s3}. The computational tools embedded within the EvtGen package enable us to compute the numerical value of the branching ratio for the $\bar B_d \to \mu^+ \mu^- e^+ e^- $ decay.

The total amplitude of the decay $\bar B_d \to \mu^+ \mu^- e^+ e^- $  can be presented in the form:
\begin{eqnarray}
\label{Mtot}
{{\cal M}^{(tot)}} = {\cal M}_{fi}^{(e )} + {\cal M}_{fi}^{( \mu)}+ {\cal M}_{fi}^{(1234)}  + {\cal M}_{fi}^{(WA)} = \nonumber\\ \,\,\,\,\,\,\,\,\,\,\,\,\,\,\,\,\,\,\,\,\,\,\,\,\,\,\,\,\,\, \sqrt{2}\, G_{F}\,\alpha^2_{em}\, V_{tb}\, V_{td}^*\,\sum\limits_{L}L\,j_1\,j_2,
\nonumber
\end{eqnarray}
where $j_1$ and $j_2$ are lepton currents, $L$ are the Lorentz structures contributed to this matrix element.

The differential branching ratio of the decay $\bar B_d \to \mu^+ \mu^- e^+ e^- $ has the form
\begin{eqnarray}
\label{dBr1234common}
d \textrm{Br} \left (\bar B_d \to \mu^+ \mu^- e^+ e^- \right ) \, =\, \tau_{B_d}\,
\frac{\sum\limits_{s_1,\, s_2,\, s_3,\, s_4}\left | {\cal M}_{fi}^{(tot)} \right |^2}{2 M_1}\, d\Phi_4 = \nonumber\\
\,\,\,\,\,\,\,\,\,\,\,\,\,\,\,\,\,\,\,\,\,\,\,\,\,\,\,\,\,\,
= \frac{G^2_{F}\,\alpha^4_{em}\, |V_{tb}\, V_{td}^*|^2}{M_1}\,\tau_{B_d}\,\sum\limits_{s_1,\, s_2,\, s_3,\, s_4}\left |\sum\limits_{L}L\,j_1\,j_2\right |^2\,d\Phi_4, 
\nonumber
\end{eqnarray}
here $\tau_{B_d}$ represents the lifetime of the $B_d$  -- meson, and the four-particle phase space $d\Phi_4$ is determined by (\ref{dPhi1234}). The summation is carried out over the spins of the final leptons $s_1, s_2, s_3,$ and $s_4$. Full integration can only be achieved numerically.

We employed the built-in capabilities of the EvtGen package to implement a multidimensional integrator using the effective geometry Monte Carlo algorithm.

Upon applying numerical integration to the $\bar B_d\to \mu^+ \mu^- e^+ e^-$ decay, the result is as follows: \begin{eqnarray}
\label{BrBdfromEvtWithPhi}
{\textrm{Br}_{1}\,\left (\bar B_d \to \mu^+ \mu^- e^+ e^- \right ) \approx (3.2\pm 1.2)*10^{-11}}.
\end{eqnarray}
This value is derived by accounting for the contributions from $\rho(770)$ and $\omega(782)$ resonances. The contribution of the $J/\psi$ resonance is  omitted from the calculation of the branching ratio $\textrm{Br}\,\left (\bar B_d \to \mu^+ \mu^- e^+ e^- \right )$ in accordance with the experimental procedure outlined. It's important to emphasize that the influence of the “tails”  from the $J/\psi$ resonance remains significant and could potentially elevate the branching ratio of $\bar B_d \to \mu^+ \mu^- e^+ e^-$. We also exclude the contribution from the $\psi(2S)$ resonance.  

To assess the significance of the contributions from higher $c\bar c$ resonances and the “tails” of the $J/\psi$ and $\psi(2S)$ resonances, the contribution of the $\omega(782)$ resonance was also excluded using the condition ${|\sqrt{M^2_1x_{ij}} - m(\omega)|} < 70\,\textrm{MeV}$. The contribution of the broad $\rho(770)$ resonance was not excluded.
The value of the partial width for the decay $\bar B_d \to \mu^+ \mu^- e^+ e^-$ is:
\begin{eqnarray}
\label{BrBdfromEvtWithoutPhi}
{\textrm{Br}_2\,\left (\bar B_d \to \mu^+ \mu^- e^+ e^- \right ) \approx (1.1\pm 0.3)*10^{-11}}.
\end{eqnarray}
Comparing  (\ref{BrBdfromEvtWithPhi}) and (\ref{BrBdfromEvtWithoutPhi}), we can observe that the contribution of the $\omega(782)$ resonance to the amplitude of the $\bar B_d \to \mu^+ \mu^- e^+ e^-$ decay is significant.
Subtraction methods for the resonant contribution in rare four-lepton $B$ decays were discussed in \cite{MeIn}. We can apply a similar subtraction method in our case, as suggested in \cite{MeIn}. The resonant contribution to the decay amplitude it is only necessary to multiply it by the dimensionless factor $q^2/M_{2i}^2$ (\ref{sec;ABC}). This multiplication also eliminates the pole dependence as $q^2$ approaches zero for the resonant contribution to the amplitude. For the case when $\omega(782)$ resonance is taken into account we obtain:
\begin{eqnarray}
\label{BrBdfromEvtWithPhi_sub}
{\textrm{Br}_{sub\,1}\,\left (\bar B_d \to \mu^+ \mu^- e^+ e^- \right ) \approx (3.1\pm 1.2)*10^{-11}}.
\end{eqnarray}
If we exclude the contribution of the $\omega(782)$ resonance we have:
\begin{eqnarray}
\label{BrBdfromEvtWithoutPhi_sub}
{\textrm{Br}_{sub\,2}\,\left (\bar B_d \to \mu^+ \mu^- e^+ e^- \right ) \approx (0.9\pm 0.2)*10^{-11}}.
\end{eqnarray}

Currently, the decay $\bar B_{d} \to \mu^+\mu^-e^+e^-$ has not been experimentally observed. However, the LHCb collaboration has set an upper limit on the partial width of the decay $\bar B_d \to \mu^+ \mu^- \mu^+ \mu^-$. This decay is described by a set of diagrams similar to the $\bar B_{d} \to \mu^+\mu^-e^+e^-$ decay, with the difference being that their number increases due to Fermi anti-symmetrization effects. The numerical values of the partial widths for these decays, $\bar B_{d} \to \mu^+\mu^-e^+e^-$ and $\bar B_d \to \mu^+ \mu^- \mu^+ \mu^-$, should also be of the same order of magnitude. The upper limit on the partial width of the $\bar B_d \to \mu^+ \mu^- \mu^+ \mu^-$ decay has been established by the LHCb collaboration.
\begin{eqnarray}
\label{BrBdUpLim}
{\textrm{Br}_{exp}\,\left (\bar B_d \to \mu^+ \mu^- \mu^+ \mu^- \right ) < 1.8*10^{-10}}.
\end{eqnarray}
The experimental upper limit (\ref{BrBdUpLim}) does not contradict with the theoretical predictions for the partial width of the $\bar B_d \to \mu^+ \mu^- e^+ e^-$ decay obtained in this study.

 In our matrix element the relative phases between all resonances is set to zero. The parametrisation of the form factors are chosen as follows: if the form factors have the pole parametrisation, the corresponding masses are taken from the PDG. If the form factor is taken from other works, the constants are taken in the form in which they are given in these works.

\section{Differential distributions}

We study the set of differential characteristics for the $\bar B_d \to \mu^+ \mu^- e^+ e^-$ decay, that demonstrate the features of this decay. We take into account contributions of $\rho^0(770)$, $\omega(782)$, $\psi(3770)$, $\psi(4040)$, $\psi(4160)$, $\psi(4415)$ -- resonances and “tails” from the  $ {J/\psi}$ and ${\psi(2S)}$ resonances, bremsstrahlung and weak annihilation.

\begin{figure}[h!]
\begin{minipage}[h]{0.47\linewidth}
\center{\includegraphics[width=1\linewidth]{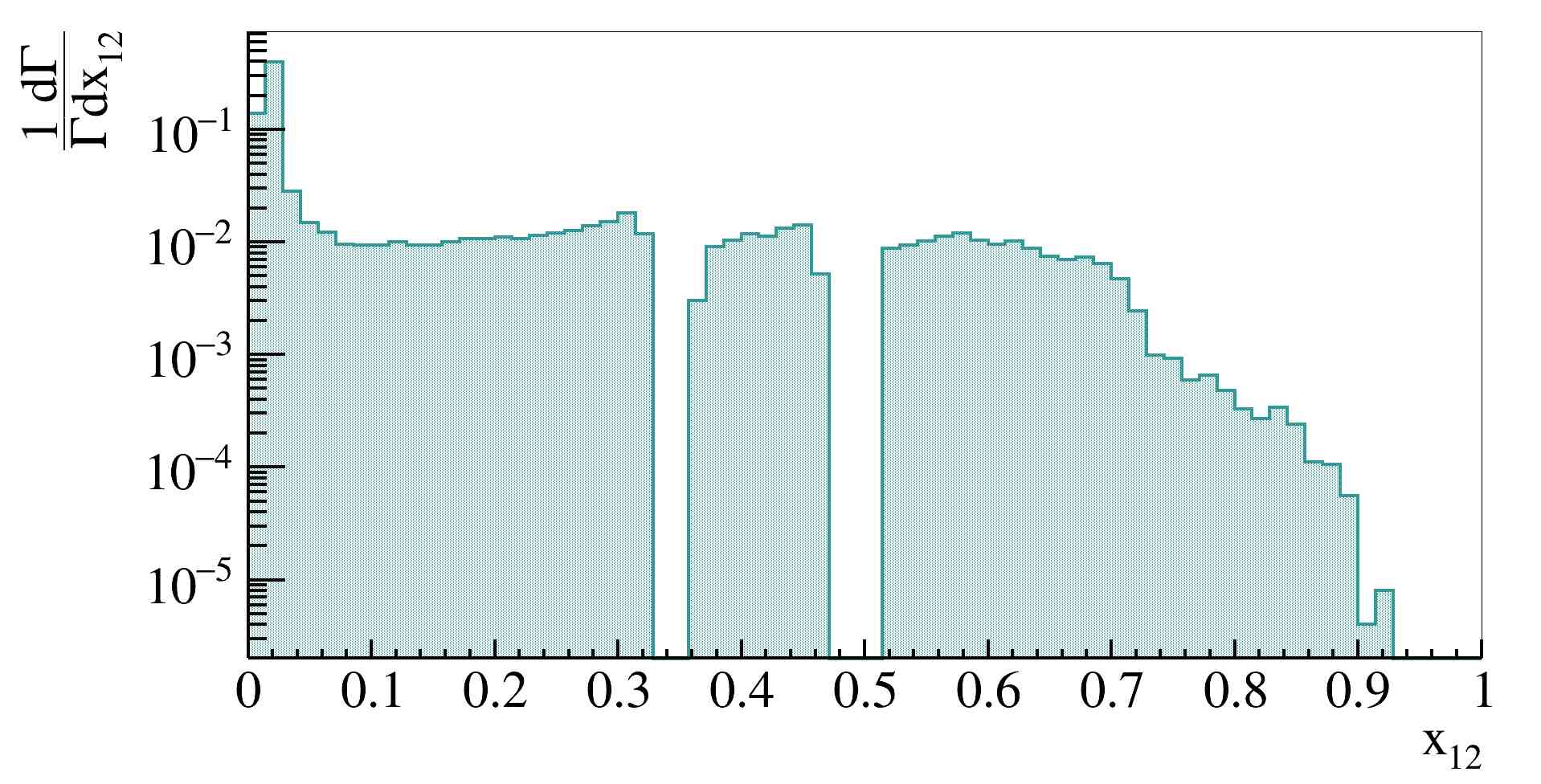}} 1) \\
\end{minipage}
\hfill
\begin{minipage}[h]{0.47\linewidth}
\center{\includegraphics[width=1\linewidth]{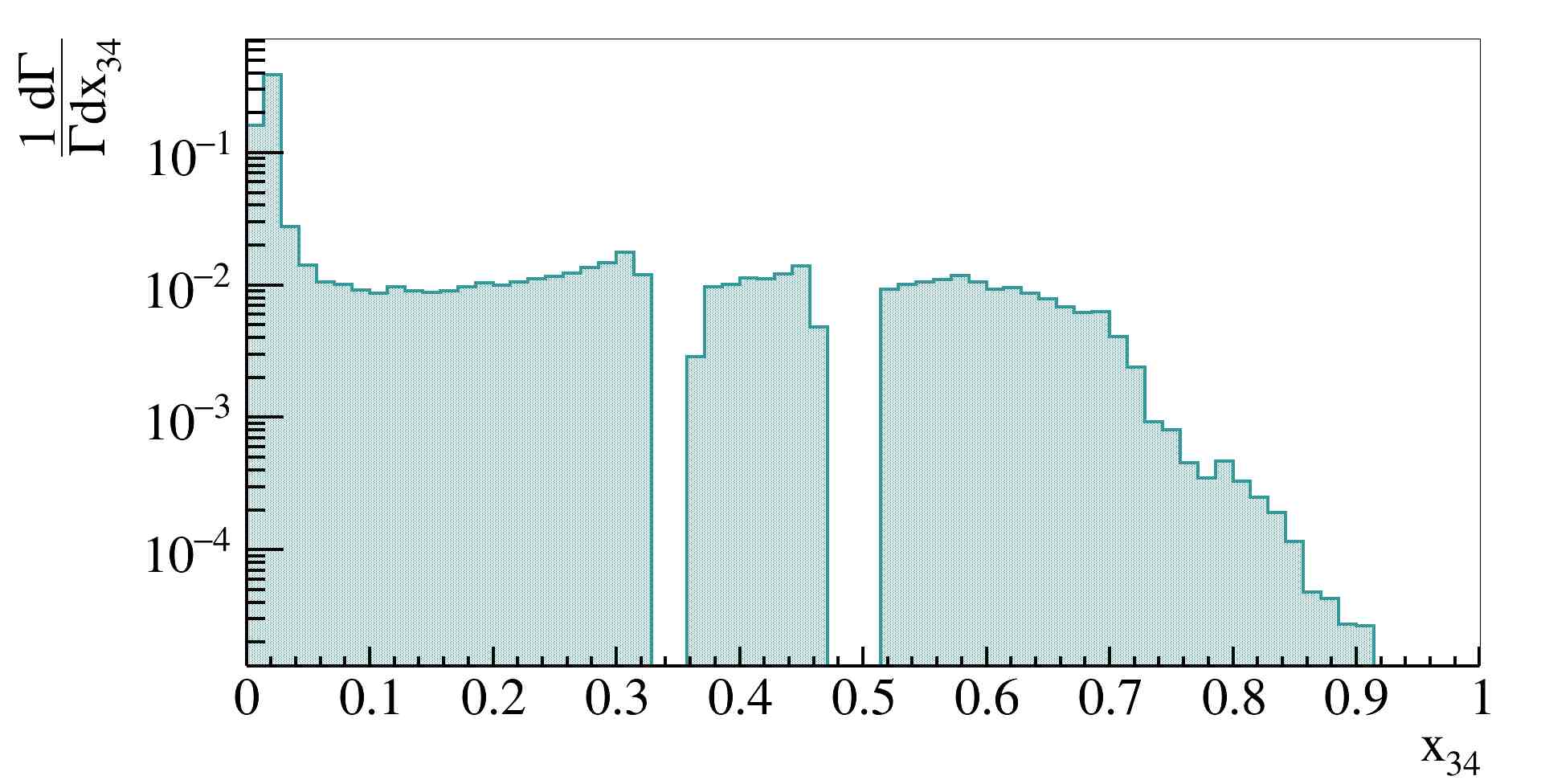}} 2)\\
\end{minipage}
\vfill
\begin{minipage}[h]{0.47\linewidth}
\center{\includegraphics[width=1\linewidth]{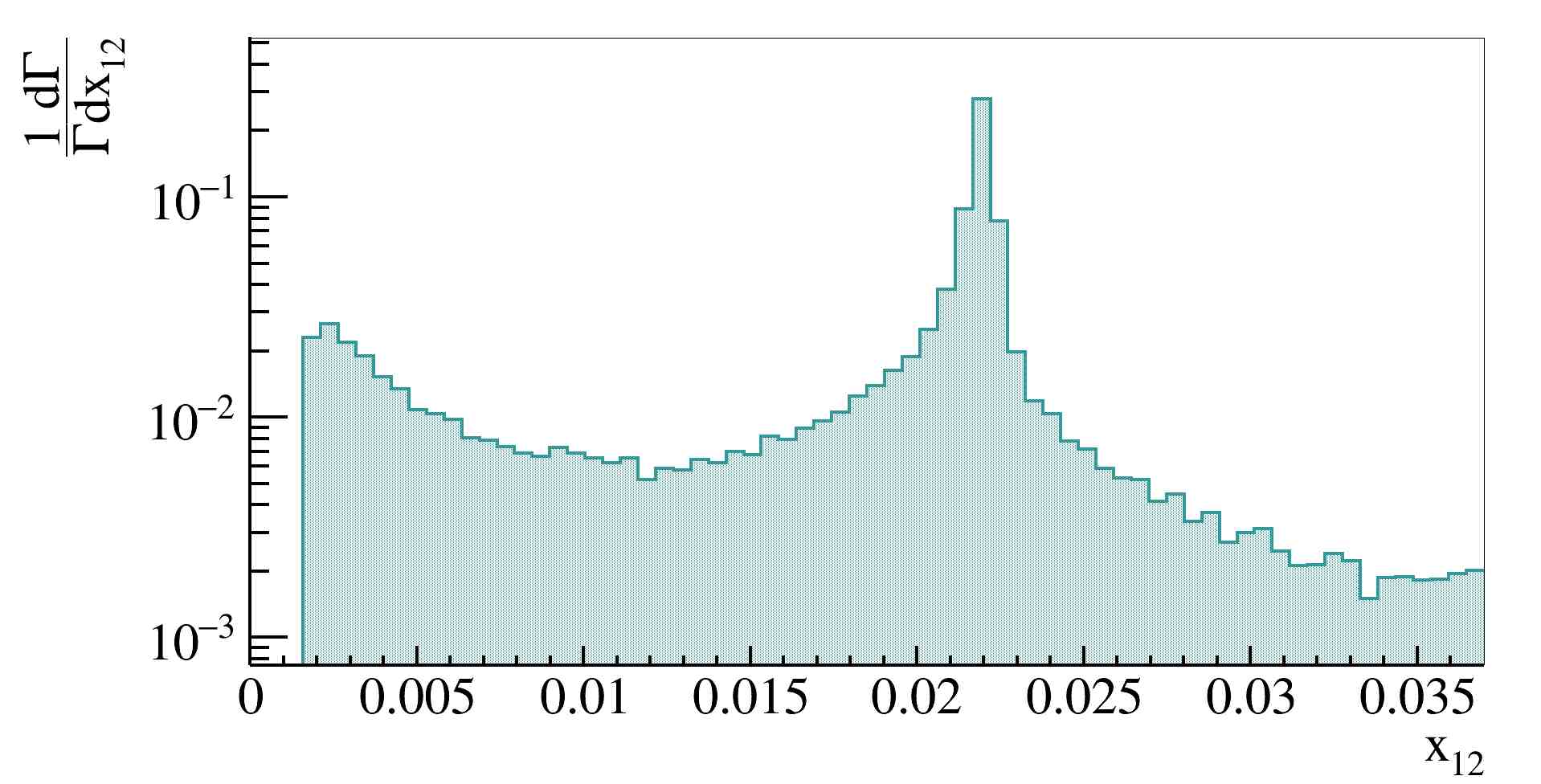}} 3) \\
\end{minipage}
\hfill
\begin{minipage}[h]{0.47\linewidth}
\center{\includegraphics[width=1\linewidth]{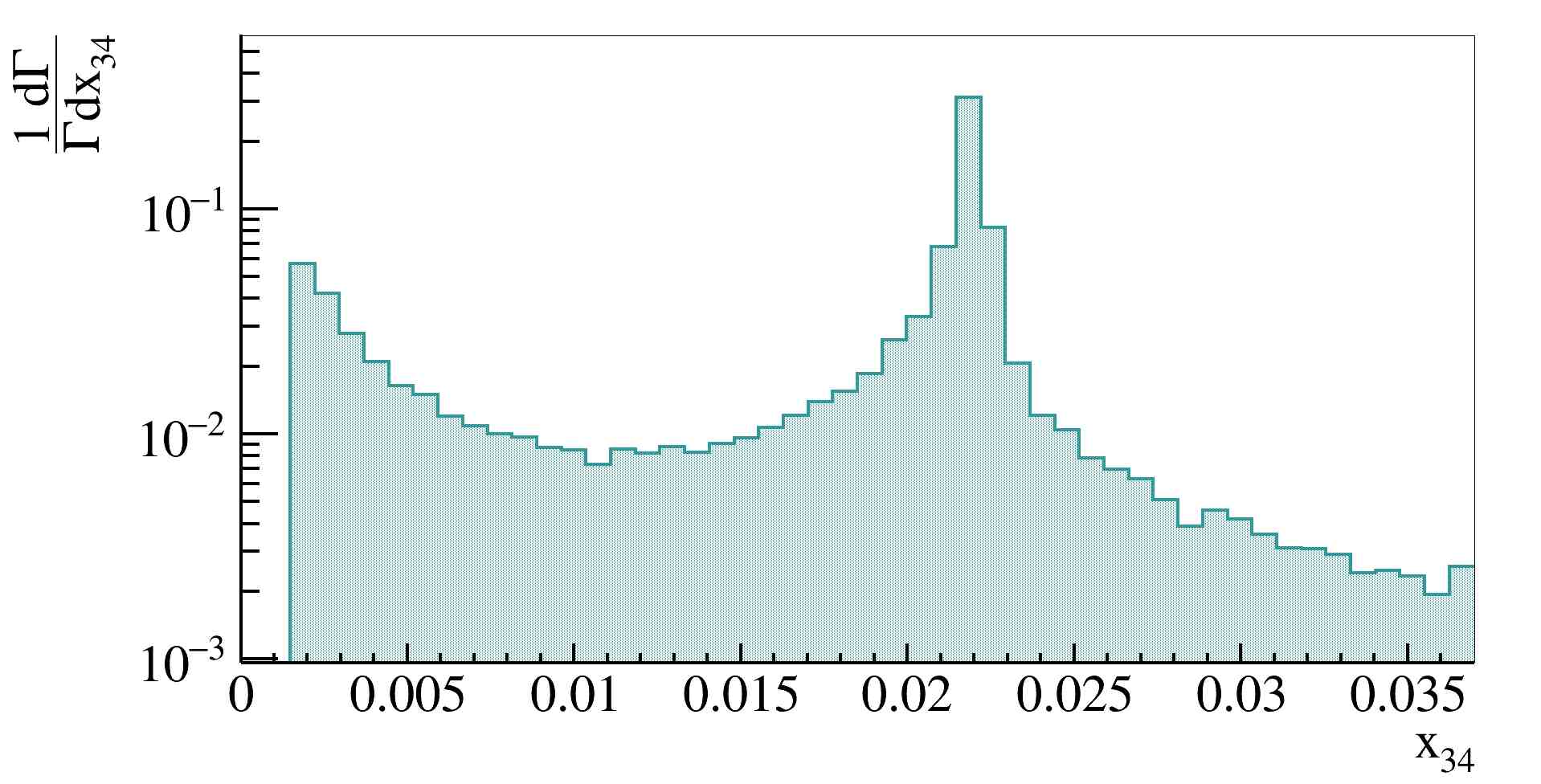}} 4)\\
\end{minipage}
\vfill
\caption {
1) Distribution of the partial decay width with respect to the variable $x_{12}$ ($\mu^+\mu^-$ channel).
2) Distribution of the partial decay width with respect to the variable $x_{34}$ ($e^+e^-$ channel).
3) Distribution of the partial decay width with respect to the variable $x_{12}$ in the region of $\rho^0(770)$ and $\omega(782)$ resonances (up to $1\,GeV^2$). 
4) Distribution of the partial decay width with respect to the variable $x_{34}$ in the region of $\rho^0(770)$ and $\omega(782)$ resonances (up to $1\,GeV^2$).}
\label {DistrBd}
\end{figure}

The distributions of the partial decay width for $\bar B_d \to \mu^+ \mu^- e^+ e^-$ as functions of $x_{12}$ and $x_{34}$ are shown in Fig.~\ref{DistrBd}. The distributions are plotted in a logarithmic scale within the range $[0.0016, 1.0]$,  as depicted in Fig.~\ref{DistrBd}. The contribution of the photon pole in the $e^+ e^-$ channel at $x_{ij\,\textrm{min}} = \frac{4 m^2_{\mu}}{M^2_1} = 0.0016$ is no significant. The shapes of the distributions for the electron and muon channels at small values of $x_{ij}$ should be similar, as demonstrated in Fig.~\ref{DistrBd}. Around $x_{ij} \to (\frac{M_{\omega}}{M_{1}})^2 \approx 0.022$, a peak from the $\omega (782)$ meson is observed. The contribution from the $\rho^0(770)$ resonance appears as a broad pedestal beneath the narrow peak from the $\omega (782)$ meson. In Fig.~\ref{DistrBd}(3, 4), the regions of $\rho^0(770)$ and $\omega (782)$ resonances are examined in more detail, with a focus on the range up to $1\,GeV^2$.

\begin{figure}[t]
\begin{center}
\begin{tabular}{cc}
\includegraphics[width=7.75cm]{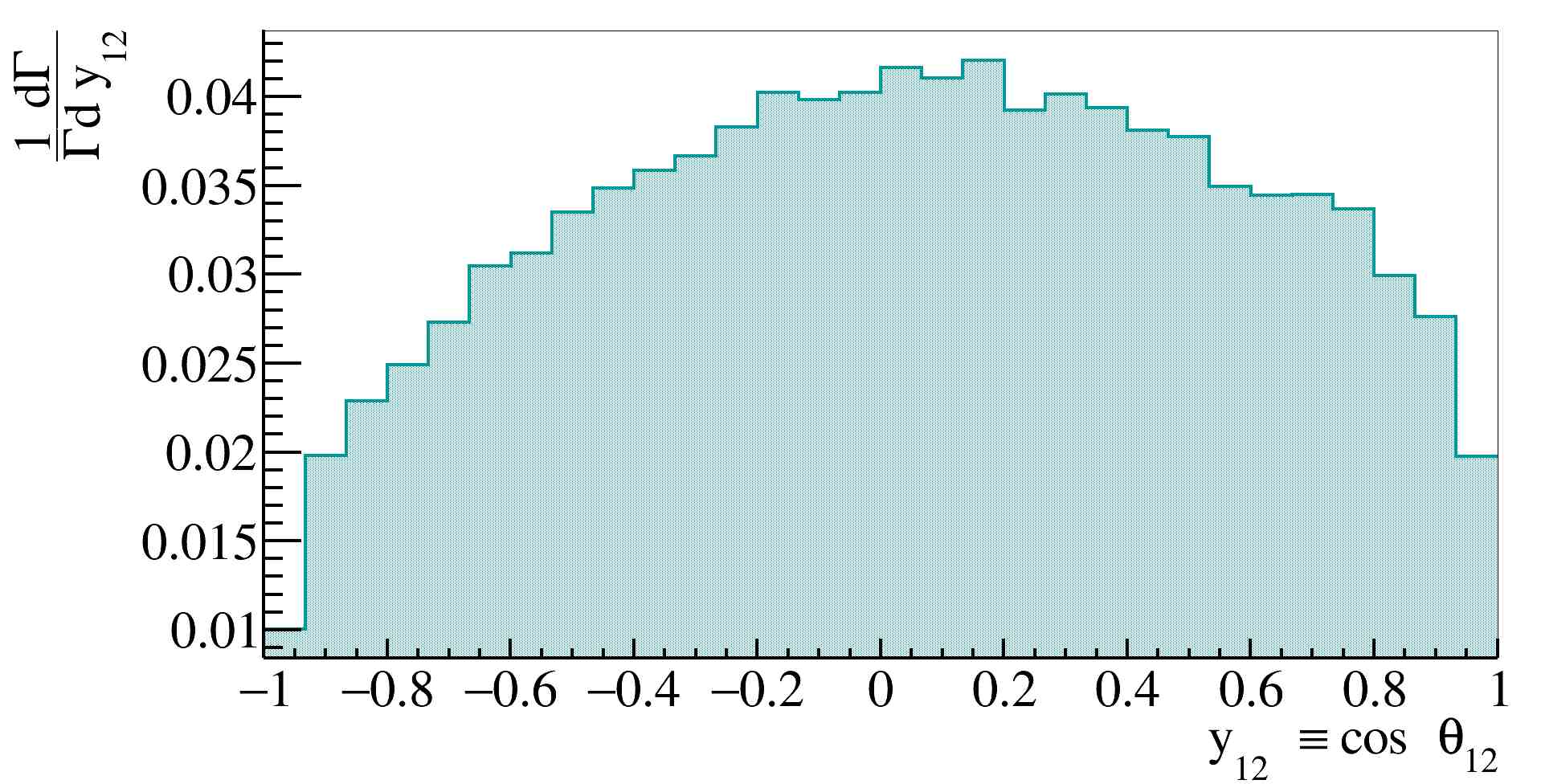} &
\includegraphics[width=7.75cm]{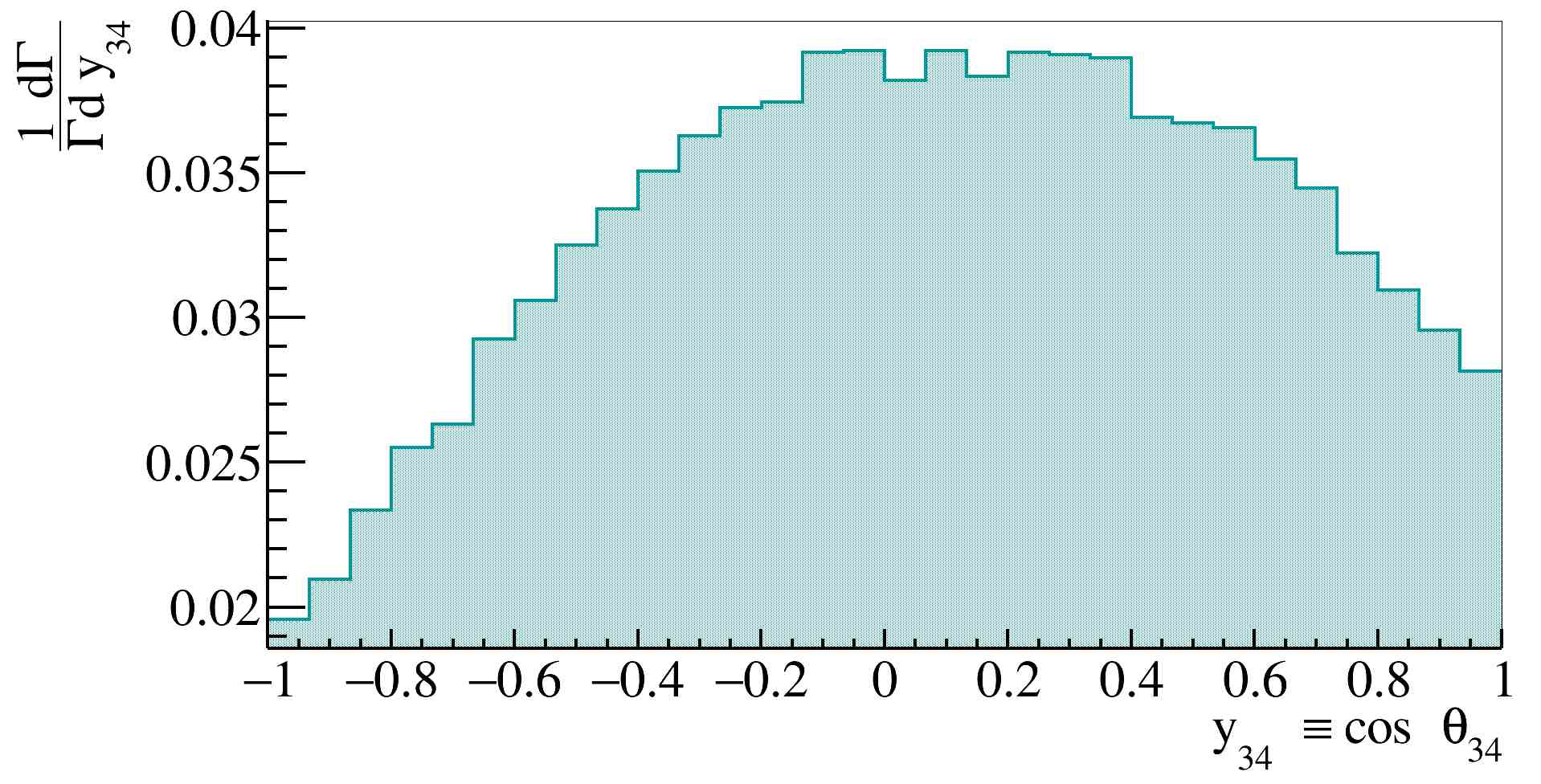} \\
\textbf{1)} & \textbf{2)} \\
\end{tabular}
\end{center}
\caption{\label{DistrBdcos} 1) Distribution of the partial decay width of $\bar B_d \to \mu^+ \mu^- e^+ e^-$ with respect to the variable $\cos(\theta_{12})$.
2) Distribution of the partial decay width of $\bar B_d \to \mu^+ \mu^- e^+ e^-$ with respect to the variable $\cos(\theta_{34})$.}
\end{figure}

\begin{figure}[t]
\begin{minipage}[h]{1.0\linewidth}
\center{\includegraphics[width=12cm]{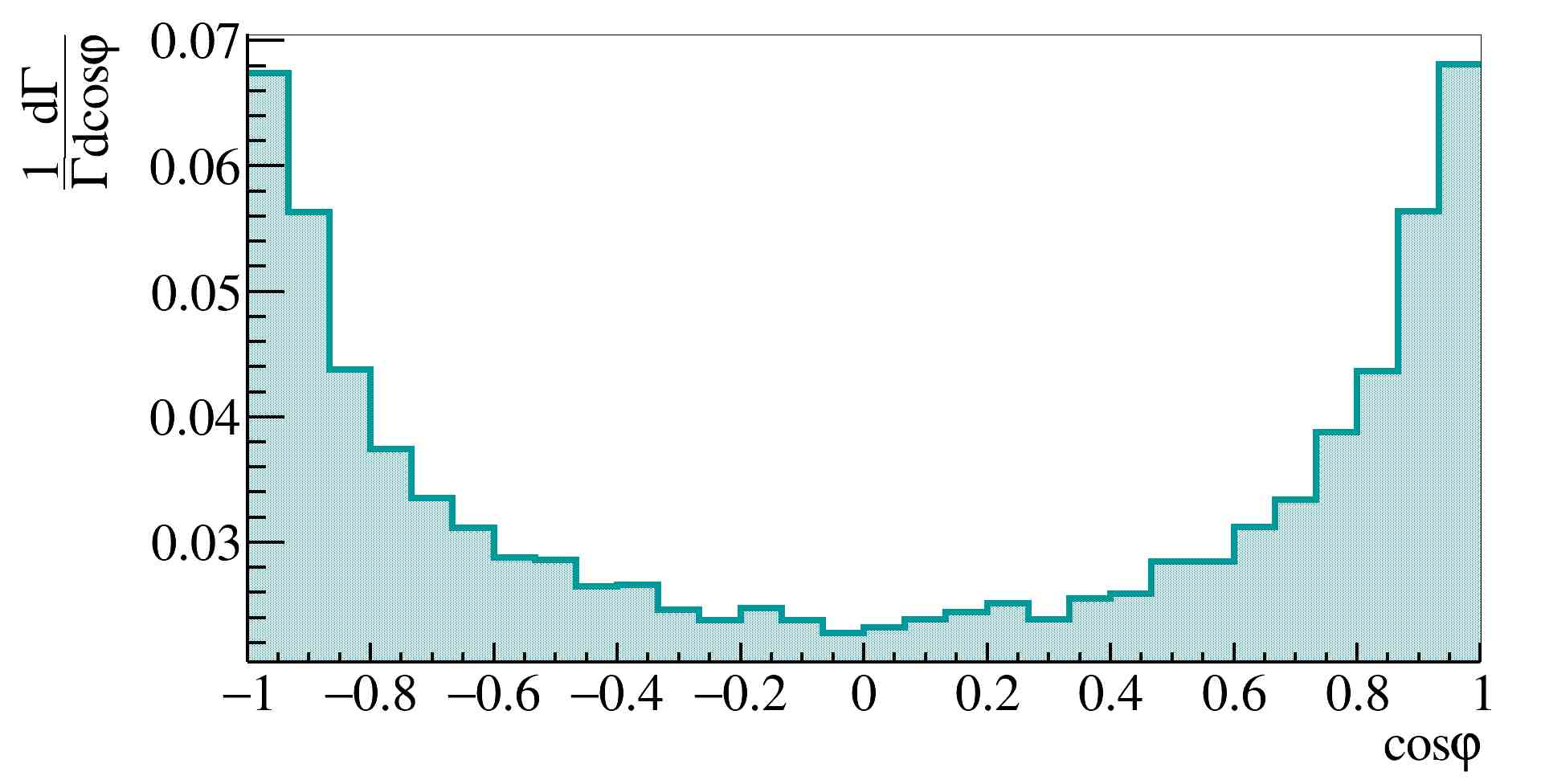}} \\
\end{minipage}
\caption{Normalized differential distribution $\displaystyle \frac{1}{\Gamma}\,\frac{d \Gamma}{d \cos(\varphi)}$.}
\label{ris:BdCosPhi}
\end{figure}

The distributions of the angular variables $y_{12} = \cos\theta_{12}$ and $y_{34} = \cos\theta_{34}$ are depicted in Fig.~\ref{DistrBdcos}(1) and Fig.~\ref{DistrBdcos}(2) respectively. Here, $\theta_{12}$ represents the angle between the propagation directions of $\mu^+$ and $B_d$ in the rest frame of the $\mu^+\mu^-$ pair. Similarly, $\theta_{34}$ is the angle between the propagation directions of $e^+$ and $B_d$ in the rest frame of the $e^+e^-$ pair.

The distribution of the partial width of the decay $\bar B_d \to \mu^+ \mu^- e^+ e^- $ with respect to the variable $\cos(\varphi)$ is shown in Fig.~\ref{ris:BdCosPhi}, where $\varphi$ is the angle between the planes of flight of the final-state lepton pairs. More precisely, $\varphi$ is defined in ~\ref{sec;kinemat4}.

The forward-backward lepton asymmetries play a crucial role in exploring physics beyond the Standard Model. These asymmetries provide valuable insights into the characteristics of the $B_d$ amplitude and the influence of intermediate resonances on the decay process.
In the context of the $\bar B_d \to \mu^+ \mu^- e^+ e^-$ decay, forward-backward lepton asymmetries $A^{(\bar{B_d})}_{FB}(x_{12})$ and $A^{(\bar{B_d})}_{FB}(x_{34})$ can be defined as
$$
\label{Afbq2-def}
A^{(\bar{B_d})}_{FB}(x_{12})\, =\,
\frac{\int\limits_0^1
d y_{12}\,\frac{\displaystyle d^2\,\Gamma \left (\bar B_d \to \mu^+ \mu^- e^+ e^- \right )}{\displaystyle d x_{12}\, d y_{12}}\,-\,
\int\limits_{-1}^0
d y_{12}\,\frac{\displaystyle d^2\,\Gamma \left (\bar B_d \to \mu^+ \mu^- e^+ e^- \right )}{\displaystyle d x_{12}\, d y_{12}}
}
{ \frac{\displaystyle d\,\Gamma \left (\bar B_d \to \mu^+ \mu^- e^+ e^- \right )}{\displaystyle d x_{12}}}
$$
and
$$
\label{Afbk2-def}
A^{(\bar{B_d})}_{FB}(x_{34})\, =\,
\frac{\int\limits_0^1
d y_{34}\,\frac{\displaystyle d^2\,\Gamma \left (\bar B_d \to \mu^+ \mu^- e^+ e^- \right )}{\displaystyle d x_{34}\, d y_{34}}\,-\,
\int\limits_{-1}^0
d y_{34}\,\frac{\displaystyle d^2\,\Gamma \left (\bar B_d \to \mu^+ \mu^- e^+ e^- \right )}{\displaystyle d x_{34}\, d y_{34}}
}
{ \frac{\displaystyle d\,\Gamma \left (\bar B_d \to \mu^+ \mu^- e^+ e^- \right )}{\displaystyle d x_{34}}}.
$$

\begin{figure}[h!]
\begin{minipage}[h]{0.47\linewidth}
\center{\includegraphics[width=1.2\linewidth]{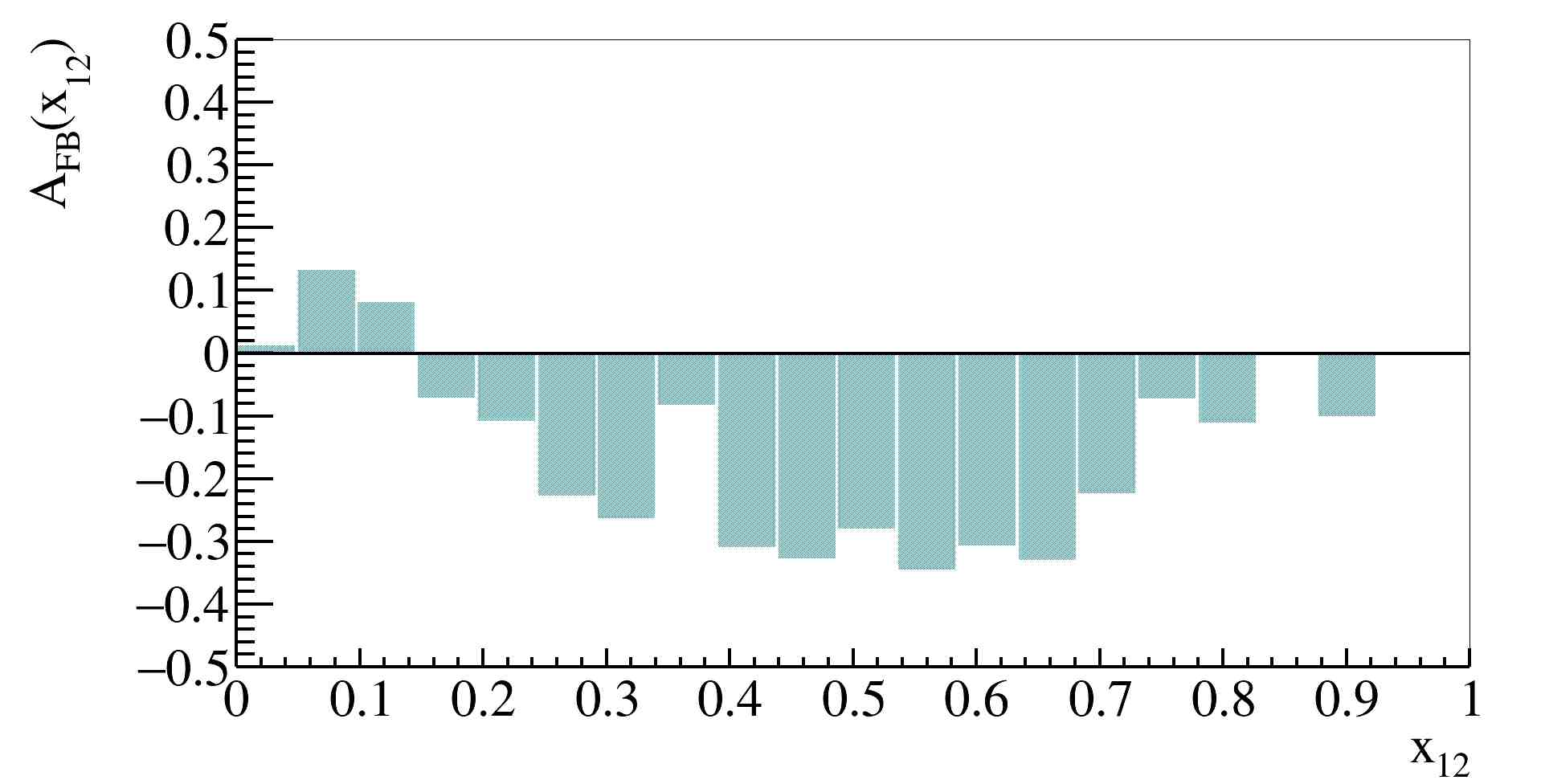}} 1) \\
\end{minipage}
\hfill
\begin{minipage}[h]{0.47\linewidth}
\center{\includegraphics[width=1.2\linewidth]{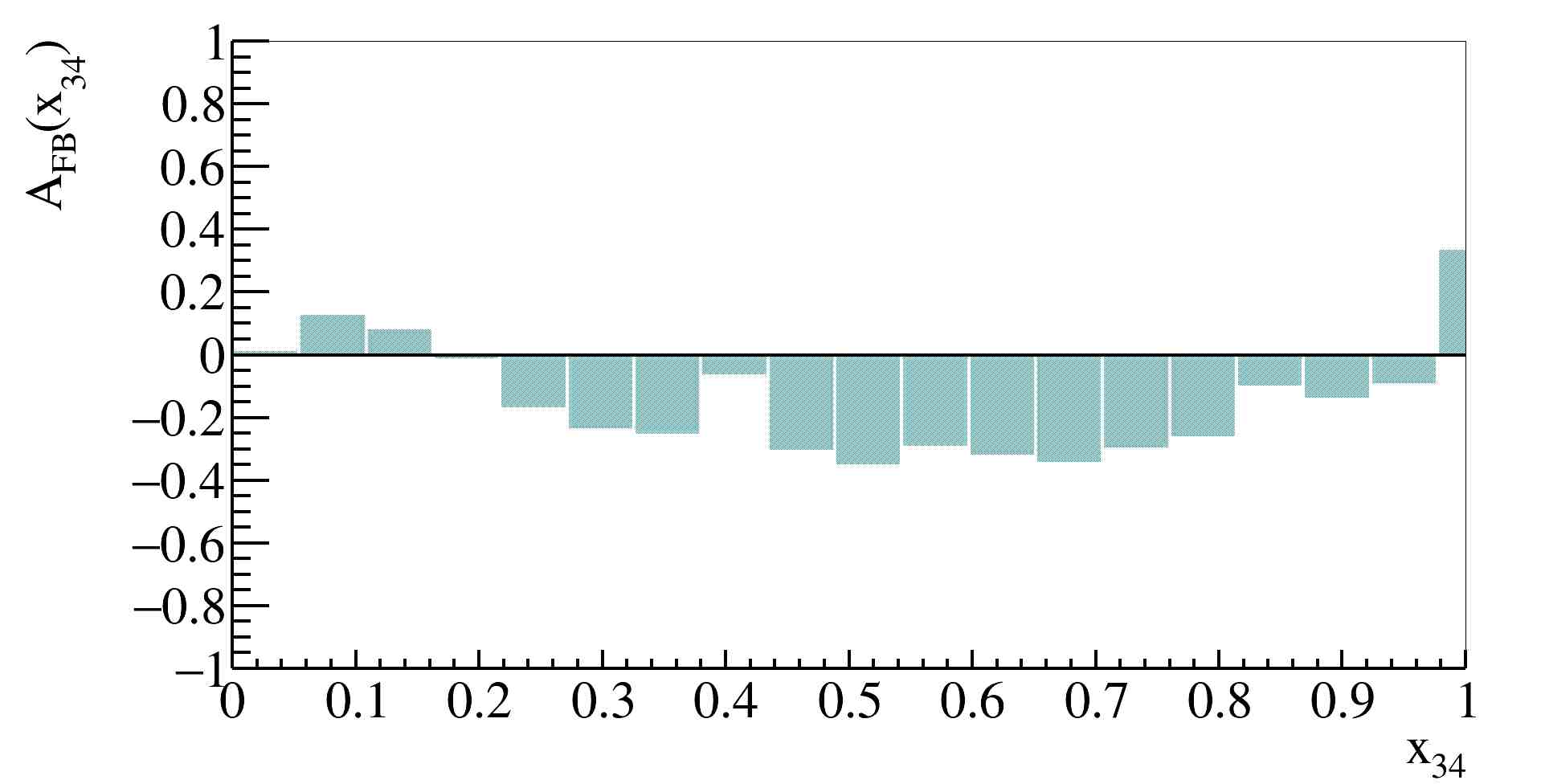}} 2)\\
\end{minipage}
\vfill
\caption{1)The forward–backward leptonic asymmetry for the $\mu^+\mu^-$ - pair; \\ 2) The forward–backward leptonic asymmetry for the $e^+e^-$ - pair.}
\label {ris:FBA}
\end{figure}

The forward-backward lepton asymmetries are depicted in Fig.~\ref{ris:FBA}. The observed shapes of these asymmetries exhibit meaningful patterns in both the $\mu^+\mu^-$ and $e^+e^-$ channels. The asymmetry values transition through zero in the region of small $x_{12}$ (or $x_{34}$), showing the influence of the $\rho^0(770)$ resonance.

In the region of larger $x_{12}$ (or $x_{34}$), the asymmetry shapes provide insights into the relative signs between the contributions of higher excited charmonium states. This behavior is particularly informative, as it sheds light on the interplay of different intermediate states and resonant structures within the decay process. Such asymmetry patterns serve as a sensitive probe to unveil subtle effects originating from specific resonance contributions and potential new physics phenomena.

\section{Conclusion}
\begin{itemize}
 \item Within the context of the Standard Model we provide predictions for the branching ratio of the decay $\bar B_d \to \mu^+ \mu^- e^+ e^- $. Our theoretical analysis encompasses various contributions, including resonant effects from $\rho(770)$ and $\omega(782)$, $\psi(3770)$, $\psi(4040)$, $\psi(4160)$, and $\psi(4415)$ resonances. Additionally, we consider the “tails” of the $J/\psi$ and $\psi(2S)$ resonances, non-resonant contributions originating from a virtual photon emitted by a $b$ -- quark, the weak annihilation process and bremsstrahlung. By excluding the contributions of $J/\psi$ and $\psi(2S)$ resonances as prescribed in~\cite{Aaij:2013lla, Aaij:2016kfs}, we obtain:
$$
{\textrm{Br}\,\left (\bar B_d \to \mu^+ \mu^- e^+ e^- \right ) \approx (3.2\pm 1.2)*10^{-10}}.
$$
Similarly, if we further exclude the effects of $\omega(782)$, $J/\psi$, and $\psi(2S)$ resonances in accordance with the methodology outlined in~\cite{4mu2021}, we obtain
$$
{\textrm{Br}\,\left (\bar B_d \to \mu^+ \mu^- e^+ e^- \right ) \approx (1.1\pm 0.3)*10^{-10}}.
$$
Comparing the branching ratios obtained with and without the presence of the $\omega(782)$ resonance reveals that the contribution from the $\omega(782)$ resonance plays a pivotal role in determining the decay amplitude.

\item Utilizing the EvtGen-based generator model, we calculate a comprehensive set of differential distributions for the $\bar B_d \to \mu^+ \mu^- e^+ e^-$ decay.
\end{itemize}


\section*{Acknowledgements}

The authors express their gratitude to  E.~E.~Boos (SINP MSU), S.~A.~Baranov(LPI), L.~V.~Dudko (SINP MSU), V.~Yu.~Yegorychev (ITEP) and A. A. Buzina(IPRAS) for engaging discussions that greatly enhanced the quality of this study.

This work was made possible with the support of Grant No. 22-22-00297 from the Russian Science Foundation (A. D. and N. N.). The authors extend their heartfelt appreciation for this valuable support.

\appendix
\section{Kinematics of four-lepton decays}
\label{sec;kinemat4}

Denote the four-momenta of  the final leptons in four-leptonic decays
of $B$--mesons as $k_i$, $i =\{1,\, 2,\, 3,\, 4\}$.  We define
$$
q = k_1 + k_2; \quad
k = k_3 + k_4; \quad
\tilde q = k_1 + k_4; \quad
\tilde k = k_2 + k_3; \quad
p = k_1 + k_2 + k_3 + k_4,
$$ 
where $p$ is the four-momentum of the $B$--meson and $p^2 = M^2_1$. For the
calculations below it is suitable to use the dimensionless
variables: 
$$
x_{12} = \frac{q^2}{M_1^2}, \quad
x_{34} = \frac{k^2}{M_1^2}, \quad
x_{14} = \frac{{\tilde q}^2}{M_1^2}, \quad
x_{23} = \frac{{\tilde k}^2}{M_1^2}.
$$
In common notation, $x_{ij} = (k_i + k_j)^2/M^2_1$. Hence $x_{ij} =
x_{ji}$. The leptons may be considered as massless in almost all of
the calculations of the present work, i.e., $k_i^2=0$. However, during
the calculation of the bremsstrahlung contribution in the area $q^2$ and $k^2$ it is necessary to take into account
the dependence of the bremsstrahlung matrix element and phase space on values of $m_e$ and $m_\mu$.

Let us find the intervals for $x_{ij}$ using the inequality
$(p_1 p_2) \ge \sqrt{p_1^2\, p_2^2}$; then any $x_{ij} \ge (\hat{m_i} + \hat{m_j})^2$. On the
other hand,
\begin{eqnarray}
1 & =&\frac{p^2}{M_1^2}\, =\,\frac{(q+k)^2}{M^2_1}\,\ge \,\frac{(\sqrt{q^2} + \sqrt{k^2})^2}{M_1^2}\, =\,\Big (\sqrt{x_{12}} + \sqrt{x_{34}} \Big)^2.
\nonumber
\end{eqnarray}
As $(\hat{m_3} + \hat{m_4})^2 \le x_{34}$, then  $x_{12} \le 1$, so $x_{12} \in [0,\, 1]$. The
upper limit of the variable $x_{34}$ depends on the value of $x_{12}$:
$$
x_{34}\, =\,\frac{(p - q)^2}{M_1^2}\,\le\,\frac{(M_1 - \sqrt{q^2})^2}{M_1^2}\, =\,\left ( 1 - \sqrt{x_{12}}\right )^2.
$$
Thus for a fixed value of $x_{12}$ the variable $x_{34} \in \left [ (\hat{m_3} + \hat{m_4})^2,\,\left ( 1 - \sqrt{x_{12}}\right )^2 \right ]$.

\begin{figure}[h!]
\begin{center}
\includegraphics[width=11.0cm]{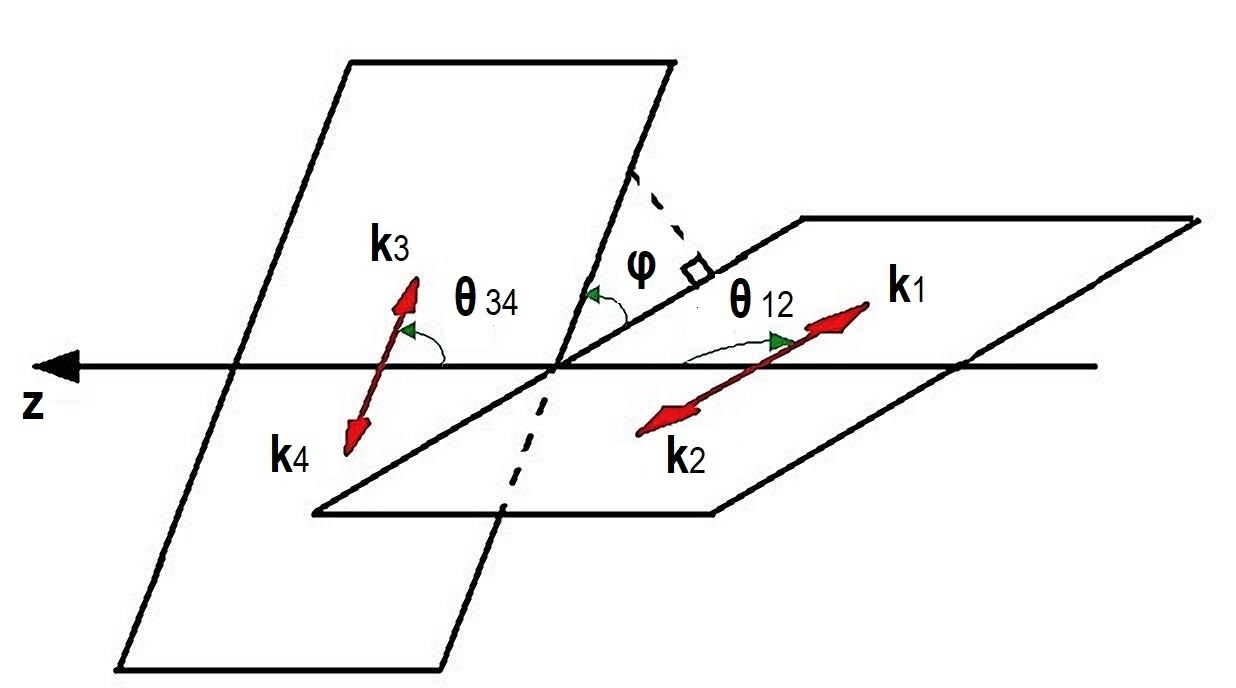} 
\end{center}
\caption{\protect\label{fig:cinematic} 
Kinematics of the decay $\bar B_d(p) \,\to\, \mu^+(k_1)\,\mu^-(k_2)\,  e^+(k_3)\,e^-(k_4)$.  Angle $\theta_{12}$ is
defined in the rest frame of $\mu^+\,\mu^-$--pair;  angle
$\theta_{34}$ is defined in the rest frame of $e^+\,e^-$--pair; angle $\varphi$ is defined
in the rest frame of $\bar{B_d}$ -- meson.
}
\end{figure}

Let us consider the kinematics of the decay
$ 
\bar B_d(p) \,\to\, \mu^+(k_1)\,\mu^-(k_2)\,  e^+(k_3)\,e^-(k_4) 
$. We define an angle
$\theta_{12}$ between the momentum of $\mu^+$ and
the direction of the $\bar{B_d}$--meson ($z$--axis) in the rest frame of the
$\mu^+\mu^-$ pair, and another angle $\theta_{34}$ between the
direction of the $e^+$ and the direction of the $\bar{B_d}$ -- meson
($z$--axis) in the rest frame of $e^+ e^-$ -- pair. Then
\begin{eqnarray}
\label{costheta-equation}
y_{12}\,\equiv\,\cos\theta_{12} &=& \frac{1}{\lambda^{1/2} (1,\, x_{12},\, x_{34})}\,\left ( x_{23} + x_{24} - x_{13} - x_{14}\right ),
\\
y_{34}\,\equiv\,\cos\theta_{34} &=& \frac{1}{\lambda^{1/2} (1,\, x_{12},\, x_{34})}\,\left ( x_{14} + x_{24} - x_{13} - x_{23}\right ),
\nonumber
\end{eqnarray}
where $\lambda (a, \, b,\, c) = a^2 + b^2 + c^2 - 2ab - 2ac - 2bc$, the 
triangle function. Angles $\theta_{12} \in \left [0,\,\pi \right ]$
and $\theta_{34} \in \left [0,\,\pi \right ]$. Hence $y_{12} \in
[-1,\, 1]$ and $y_{34} \in [-1,\, 1]$. Angles are measured relative to
$z$--axis. Also let us define an angle $\varphi \in [0,\, 2 \pi)$ in
the rest frame of the $\bar{B_d}$ -- meson between the planes which are set by the pairs
of vectors $({\bf k}_1,\, {\bf k}_2)$ and $({\bf k}_3,\, {\bf
  k}_4)$.  Introduce a vector ${\bf  a}_1 = {\bf k}_1\,\times\,
{\bf k}_2$, perpendicular to the plane $({\bf k}_1,\, {\bf k}_2)$, and 
vector ${\bf  a}_3 = {\bf k}_{\, 4}\,\times\, {\bf k}_{\, 3}$, which are
normal to the plane $({\bf k}_3,\, {\bf k}_4)$. Then
$$
\cos\varphi\, =\,\frac{\Big ( {\bf  a}_1,\, {\bf  a}_3\Big )}{|{\bf  a}_1|\, |{\bf  a}_3|}.
$$

It is suitable to choose $x_{12}$, $x_{34}$, $y_{12}$, $y_{34}$, and
$\varphi$ as independent integration variables. Then the four body phase space has the form
 
\begin{eqnarray}
\label{dPhi1234}
\fl
 \,\,\,\,\,\,\,\,\,d\Phi_4=                                                    \frac{M_1^4}{2^{14}\, \pi^6}\,
\lambda^{1/2} \left ( 1,\,  x_{12},\, x_{34}\right )\,\sqrt{1\, -\,\frac{4 {\hat m}_\mu^2}{x_{12}}}\,\, \sqrt{1\, -\,\frac{4{\hat m}_{e}^2}{x_{34}}}
d x_{12}\, d x_{34}\, d y_{12}\, d y_{34}\, d \varphi, 
\end{eqnarray} 
where ${\hat m}_\mu = m_\mu /M_1$ and  ${\hat m}_{e} = m_{e}/M_1$.

This paper uses notations almost  identical
to the notations of Ref. \cite{Barker:2002ib}, except for 
$y_{ij}$, which here have the opposite sign compared to
Ref. \cite{Barker:2002ib}.

\section{The
functions $a^{(ij)}, b^{(ij)}$, $c^{(ij)}$, $d^{(ij)}$, $f^{(ij)}$ and $g^{(ij)}$}
Here we define the dimensionless functions from the (\ref{Amp}) decay amplitude.
\label{sec;ABC}
\begin{eqnarray}
\label{AVV}
\fl a^{(VV)}(x_{12}\, x_{34}) = 
\frac{1}{M_1^2}\,\Bigg[ \frac{4\,\hat m_{b}\, C_{7\gamma}(\mu)}{x_{12},\, x_{34}}\,\Bigg(\frac{1}{2}(F_{TV}(q^2,k^2)+F_{TV}(k^2,q^2)) -\nonumber\\ - \sum\limits_{i= \rho^0,\,\omega}
\,\frac{I_i\, \hat M_{2i}\, \hat f_{V_i}}{x_{34}-\hat M_{2i}^2+i\hat \Gamma_{2i} M_{2i}}{\cal F}_{\mu\nu}^{(i)}(k^2) \,T_1(q^2) -\nonumber\\ - \sum\limits_{i= \rho^0,\,\omega}
\,\frac{I_i\, \hat M_{2i}\, \hat f_{V_i}}{x_{12}-\hat M_{2i}^2+i\hat \Gamma_{2i} \hat M_{2i}}{\cal F}_{\mu\nu}^{(i)}(k^2) \,T_1(k^2)\Bigg) +\nonumber\\ +\,
\frac{C_{9V}(q^2,\,\mu)}{x_{34}}\,\Bigg(F_{V}(q^2,k^2)  - \frac{2\hat M_{2i}}{1 + \hat M_{2i}}\,\sum\limits_{i= \rho^0,\,\omega}
\,\frac{I_i\, V(q^2)\, \hat f_{V_i}}{x_{34}-\hat M_{2i}^2+i\hat\Gamma_{2i} M_{2i}}{\cal F}_{\mu\nu}^{(i)}(k^2) \,\Bigg) +\nonumber\\ +\,
\frac{C_{9V}(k^2,\,\mu)}{x_{12}}\,\Bigg(F_{V}(k^2,q^2)  - \frac{2\hat M_{2i}}{1 + \hat M_{2i}}\,\sum\limits_{i= \rho^0,\,\omega}\frac{I_i V(k^2)\,\hat f_{V_i}}{x_{12} - \hat M_{2i}^2+i\hat \Gamma_{2i} \hat M_{2i}}\,\Bigg)\Bigg]\,;
\nonumber
\end{eqnarray}
\\
\begin{eqnarray}
\label{AVA}
\fl a^{(VA)}(x_{12},\, x_{34}) = \frac{1}{M_1^2}\,\frac{C_{10A}}{x_{12}}\,\Bigg[ \,F_{V}(k^2,q^2)  - \frac{2\hat M_{2i}}{1 + \hat M_{2i}}\,\sum\limits_{i= \rho^0,\,\omega}\frac{I_i V(k^2)\,\hat f_{V_i}}{x_{12} - \hat M_{2}^2+i\hat \Gamma_{2i} \hat M_{2i}}\,\Bigg]\,;
\nonumber
\end{eqnarray}
\\
\begin{eqnarray}
\label{AAV}
\fl a^{(AV)}(x_{12},\, x_{34}) = \frac{1}{M_1^2}\,\frac{C_{10A}}{x_{34}}\,\Bigg[ \,F_{V}(q^2,k^2)  - \frac{2\hat M_{2i}}{1 + \hat M_{2i}}\,\sum\limits_{i= \rho^0,\,\omega}\frac{ I_i V(q^2)\,\hat f_{V_i}}{x_{34} - \hat M_{2i}^2+i\hat \Gamma_{2i} \hat M_{2i}}\,\Bigg]\,;
\nonumber
\end{eqnarray}
\\
\begin{eqnarray}
\fl b^{(VV)}(x_{12},\, x_{34}) = \frac{1}{M_1^2}\,\Bigg[ \frac{2\,\hat m_{b}\, C_{7\gamma}(\mu)}{x_{12},\, x_{34}}\,\Bigg(\frac{1 - x_{12} -x_{34}}{2}(F_{TA}(q^2,k^2)+F_{TA}(k^2,q^2)) -\nonumber\\ - \sum\limits_{i= \rho^0,\,\omega}\frac{\hat M_{2i}\,(1 - \hat M_{2i}^2)\,I_i\hat f_{V_i}}{x_{34} - \hat M_{2i}^2+i\hat \Gamma_{2i} \hat M_{2i}}\,T_2(q^2) - \sum\limits_{i= \rho^0,\,\omega}\frac{\hat M_{2i}\,(1 - \hat M_{2i}^2)\,I_i\hat f_{V_i}}{x_{12} - \hat M_{2i}^2+i\hat \Gamma_{2} \hat M_{2i}}\,T_2(k^2)\Bigg) +\nonumber\\ +\,
\frac{C_{9V}(q^2,\,\mu)}{x_{34}}\,\Bigg(\frac{1}{2}(1 - x_{12} -x_{34})F_{A}(q^2,k^2)  - \sum\limits_{i= \rho^0,\,\omega}\frac{\hat M_{2i}\,(1 + \hat M_{2i})\,I_i\hat f_{V_i}}{x_{34} - \hat M_{2i}^2+i\hat \Gamma_{2} \hat M_{2i}}\,A_1(q^2)\Bigg) +\nonumber\\ +\,
\frac{C_{9V}(k^2,\,\mu)}{x_{12}}\,\Bigg(\frac{1}{2}(1 - x_{12} -x_{34})F_{A}(k^2,q^2)  - \sum\limits_{i= \rho^0,\,\omega}\frac{\hat M_{2i}\,(1 + \hat M_{2i})\,I_i\hat f_{V_i}}{x_{12} - \hat M_{2i}^2+i\hat \Gamma_{2i} \hat M_{2i}}\,A_1(k^2)\Bigg)\Bigg]\,;
\nonumber
\end{eqnarray}
\\
\begin{eqnarray}
\label{BVA}
\fl b^{(VA)}(x_{12},\, x_{34}) = \frac{1}{M_1^2}\,\frac{C_{10A}(\mu)}{x_{12}}\,\Bigg[ \,\frac{1\,-x_{12}-x_{34}}{2}\,F_{A}(k^2,q^2)  - \,\sum\limits_{i= \rho^0,\,\omega}\frac{\hat M_{2i}\,(1 + \hat M_{2i})\,I_i\hat f_{V_i}}{x_{12} - \hat M_{2i}^2+i\hat \Gamma_{2i} \hat M_{2i}}\,A_1(k^2)\,\Bigg]\,;
\nonumber
\end{eqnarray}
\\
\begin{eqnarray}
\label{abc_x12x34}
\fl b^{(AV)}(x_{12},\, x_{34}) = \frac{1}{M_1^2}\,\frac{C_{10A}(\mu)}{x_{34}}\,\Bigg[ \,\frac{1\,-x_{12}-x_{34}}{2}\,F_{A}(q^2,k^2)  - \,\sum\limits_{i= \rho^0,\,\omega}\frac{\hat M_{2i}\,(1 + \hat M_{2i})\,I_i\hat f_{V_i}}{x_{34} - \hat M_{2i}^2+i\hat \Gamma_{2i} \hat M_{2i}}\,A_1(q^2)\,\Bigg]\,;
\nonumber
\end{eqnarray}
\\
\begin{eqnarray}
\fl c^{(VV)}(x_{12},\, x_{34}) = \frac{1}{M_1^2}\,\Bigg[ \frac{2\,\hat m_{b}\, C_{7\gamma}(\mu)}{x_{12},\, x_{34}}\,\Bigg(\frac{1}{2}(F_{TA}(q^2,k^2)+F_{TA}(k^2,q^2)) - \nonumber\\ - \sum\limits_{i= \rho^0,\,\omega}\frac{\hat M_{2i}\,I_i\hat f_{V_i}}{x_{34} -  \hat M_{2i}^2+i\hat \Gamma_{2i} \hat M_{2i}}\,\Bigg(T_2(q^2) + \frac{T_3(q^2)\,x_{12}}{(1 - \hat M_{2i}^2)\,}\Bigg) -\nonumber\\ - \sum\limits_{i= \rho^0,\,\omega}\frac{\hat M_{2i}\,I_i\hat f_{V_i}}{x_{12} - \hat M_{2i}^2+i\hat \Gamma_{2} \hat M_{2i}}\,\Bigg(T_2(k^2) + \frac{T_3(k^2)\,x_{34}}{(1 - \hat M_{2i}^2)\,}\Bigg)\Bigg) +\nonumber\\ +\,
\frac{C_{9V}(q^2,\,\mu)}{x_{34}}\,\Bigg(\frac{1}{2}F_{A}(q^2,k^2)  - \sum\limits_{i= \rho^0,\,\omega}\frac{\hat M_{2i}}{(1 + \hat M_{2i})}\frac{A_2(q^2) I_i\hat f_{V_i}}{x_{34} - \hat M_{2i}^2+i\hat \Gamma_{2i} \hat M_{2i}}\Bigg) +\nonumber\\ +\,
\frac{C_{9V}(k^2,\,\mu)}{x_{12}}\,\Bigg(\frac{1}{2}F_{A}(k^2,q^2)  - \sum\limits_{i= \rho^0,\,\omega}\frac{\hat M_{2i}}{(1 + \hat M_{2i})}\frac{A_2(k^2) I_i\hat f_{V_i}}{x_{12} - \hat M_{2i}^2+i\hat \Gamma_{2i} \hat M_{2i}}\Bigg)\Bigg]\,;
\nonumber
\end{eqnarray}
\\
\begin{eqnarray}
\label{CVA}
\fl c^{(VA)}(x_{12},\, x_{34}) = \frac{1}{M_1^2}\,\frac{C_{10A}(\mu)}{x_{12}}\,\Bigg[ \,\frac{1}{2}\,F_{A}(k^2,q^2)  - \sum\limits_{i= \rho^0,\,\omega}\frac{\hat M_{2i}}{1 + \hat M_{2i}}\,\frac{A_2(k^2)\,I_i\hat f_{V_i}}{x_{12} - \hat M_{2i}^2+i\hat \Gamma_{2i} \hat M_{2i}}\,\Bigg]\,;
\nonumber
\end{eqnarray}
\\
\begin{eqnarray}
\label{CAV}
\fl c^{(AV)}(x_{12},\, x_{34}) = \frac{1}{M_1^2}\,\frac{C_{10A}(\mu)}{x_{34}}\,\Bigg[ \,\frac{1}{2}\,F_{A}(q^2,k^2)  - \sum\limits_{i= \rho^0,\,\omega}\frac{\hat M_{2i}}{1 + \hat M_{2i}}\,\frac{A_2(q^2)\,I_i\hat f_{V_i}}{x_{34} - \hat M_{2i}^2+i\hat \Gamma_{2i} \hat M_{2i}}\,\Bigg]\,;
\nonumber
\end{eqnarray}
\\
\begin{eqnarray}
\label{DAV}
\fl d^{(AV)}(x_{12},\, x_{34}) = \frac{1}{M_1^2}\,\frac{C_{10A}(\mu)}{x_{34}}\,\sum\limits_{i= \rho^0,\,\omega}\frac{I_i\hat f_{V_i}}{x_{34} - \hat M_{2i}^2+i\hat \Gamma_{2i} \hat M_{2i}} \,\Bigg[\frac{ A_{2}(q^2)}{1 + \hat M_{2i}  } + \frac{2\hat M_{2i}}{x_{12}}\Bigg(A_{3}(q^2) - A_{0}(q^2)\Bigg)\Bigg];
\nonumber
\end{eqnarray}
\\
\begin{eqnarray}
\label{GVA}
\fl g^{(VA)}(x_{12},\, x_{34}) = \frac{1}{M_1^2}\,\frac{C_{10A}(\mu)}{x_{12}}\,\sum\limits_{i= \rho^0,\,\omega}\frac{I_i\hat f_{V_i}}{x_{12} - \hat M_{2i}^2+i\hat \Gamma_{2} \hat M_{2i}} \,\Bigg[\frac{ A_{2}(k^2)}{1 + \hat M_{2i}  } + \frac{2\hat M_{2i}}{x_{34}}\Bigg(A_{3}(k^2) - A_{0}(k^2)\Bigg)\Bigg];\nonumber
\end{eqnarray}
\\
The dimensionless functions for the bremsstrahlung amplitude (Eq.~\ref{MuBrem}) are:
\begin{eqnarray}
\fl { d^{(VP)} (x_{12},\, x_{123},\, x_{124})\,}\, =\,{-\,\frac{4 C_{10A} \hat m_e \hat f_{B_d}}{M^2_1}\,
\frac{1}{x_{12}\, (x_{124} - \hat m_e^2)\, (x_{123} - \hat m_e^2)}\,
\frac{(k_3 - k_4,\, q)}{M^2_1}},
\nonumber
\end{eqnarray}
\\
\begin{eqnarray}
\fl {f^{(VT)} (x_{12},\, x_{123},\, x_{124})}\, =\,{-\,\frac{2 C_{10A} \hat m_e \hat f_{B_d}}{M^2_1}\,\frac{1}{x_{12}\, (x_{124} - \hat m_e^2)\, (x_{123} - \hat m_e^2)}\,
\frac{1 + x_{12}-x_{34}}{2}}.
\nonumber
\end{eqnarray}
These are $d$ -- and $f$ -- functions for the ${\mu^+\mu^-}$-- pair emitted by electron and positron in the final state (see first two diagrams on Fig. \ref{fig:Fbrem}). For the ${e^+e^-}$ -- pair emitted by $\mu^+$ and $\mu^-$ functions are similar.\\
In all formulae we use dimensionless variables $x_{12}= q^2/M_1^2$, and $x_{34} =
k^2/M_1^2$, $\hat f_{V_i} = f_{V_i}/M_1$, $\hat
M_{2i} = M_{2i}/M_1$,  and $\hat \Gamma_{2i}
= \Gamma_{2i}/M_1$; $M_{2i}$ and $\Gamma_{2i}$ are the mass and width of the $\rho(770)$ and $\omega(782)$ mesons.


Dimensionless functions
$a^{(IJ)} (x_{12},\, x_{34})$, ... ,$g^{(IJ)} (x_{12},\, x_{34})$ depend on form factors $ F_i(q_1^2, q_2^2)$, ${i\, =\,\{V,\, A,\, TV,\, TA\}}$,
$V(q^2)$,  $T_k(q^2)$, ${k\, =\,\{1,\, 2,\, 3\}}$ and $A_j(q^2)$, ${j\, =\,\{0,\, 1,\, 2,\, 3\}}$.

Form factors
$V(q^2)$, $T_1(q^2)$, $T_2(q^2)$, $T_3(q^2)$, $A_0(q^2)$, $A_1(q^2)$, $A_2(q^2)$ and $A_3(q^2)$
are taken from~\cite{MeSt}.
Form factors $ F_i(q_1^2, q_2^2)$ have been previously described in \cite{MNK, MN2004} for the decay $B_q\to l^+l^-\gamma$, when the one of the four-momenta squares was set $q_j^2 = 0$.

\end{document}